\documentclass[12pt]{article}
\usepackage{amsmath,amssymb,amsthm,amsxtra,overpic,bbm,bm,epsfig,subfigure}
\usepackage{hyperref}
\usepackage{mathrsfs}
\usepackage{graphicx}
\usepackage{multirow}
\usepackage{color}
\usepackage{comment}
\usepackage{epstopdf}
\numberwithin{equation}{section}
\usepackage{float}
\usepackage{cite}
\usepackage{hyperref}
\hypersetup{
	colorlinks=true,
	linkcolor=black,
	citecolor=black,    
	urlcolor=black,
}
\usepackage{url}
\usepackage{slashed,stmaryrd}
\usepackage{longtable}
\usepackage{extarrows}

\newcommand{\Rea}{{\rm Re}\,}

\newcommand{\tr}{{\rm Tr}}
\begin{document}

\def\thefootnote{\fnsymbol{footnote}}
\vspace{0.2cm}
\begin{center}
{\Large\bf Sphaleron in the Higgs Triplet Model}
\end{center}
\vspace{0.2cm}

\begin{center}
{\bf Jiahang Hu}~$^{a}$~\footnote{E-mail: hujiahang20@mails.ucas.ac.cn},\quad
{\bf Bingrong Yu}~$^{a,~b}$~\footnote{E-mail: yubr@ihep.ac.cn (corresponding author)},\quad
{\bf Shun Zhou}~$^{a,~b}$~\footnote{E-mail: zhoush@ihep.ac.cn (corresponding author)}
\\
\vspace{0.2cm}
{\small
$^a$ School of Physical Sciences, University of Chinese Academy of Sciences, Beijing 100049, China\\
$^b$ Institute of High Energy Physics, Chinese Academy of Sciences, Beijing 100049, China
}
\end{center}

\vspace{0.5cm}
	
\begin{abstract}
The Higgs triplet model (HTM) extends the Standard Model (SM) by one complex triplet scalar (also known as the type-II seesaw model), offering a simple and viable way to account for nonzero neutrino masses. On the other hand, the nontrivial couplings of the triplet to the gauge fields and to the SM Higgs field are expected to influence the topological vacuum structure of the SM, and consequently, the energy and the field configuration of the electroweak sphaleron. The sphaleron process plays a crucial role in dynamically generating the baryon asymmetry of the Universe. In this work, we study the vacuum structure of the gauge and Higgs fields and calculate the saddle-point sphaleron configuration in the HTM. The coupled nonlinear equations of motion of the sphaleron are solved using the spectral method. We find the inclusion of the triplet scalar could in principle significantly change the sphaleron energy compared with the SM. Nevertheless, at zero temperature, the current stringent experimental constraint on the vacuum expectation value of the triplet suppresses the difference. Interestingly, we find that there still exists some narrow parameter space where the sphaleron energy can be enhanced up to 30\% compared with the SM case.   
\end{abstract}

\newpage
	
\def\thefootnote{\arabic{footnote}}
\setcounter{footnote}{0}


\section{Introduction}
\label{sec:intro}
Despite its great success, the Standard Model (SM) of particle physics is unable to accommodate nonzero neutrino masses, which has been firmly established by the neutrino oscillation experiments during the last two decades~\cite{ParticleDataGroup:2022pth,Esteban:2020cvm} (see, e.g., Ref.~\cite{Xing:2020ijf} for a recent theoretical review). Another important unsolved problem in the SM is the observed baryon asymmetry of the Universe~\cite{Bodeker:2020ghk}. Given the 125~GeV Higgs boson discovered at the Large Hadron Collider~\cite{ATLAS:2012yve,CMS:2012qbp}, the SM cannot provide a successful electroweak (EW) baryogenesis since the EW phase transition in the SM is a smooth cross-over~\cite{DOnofrio:2014rug,DOnofrio:2015gop}, failing to depart from thermal equilibrium~\cite{Sakharov:1967dj}. Therefore, the SM should be incomplete, and new physics beyond the SM is indispensable.

The extension of the SM by adding one triplet scalar with hypercharge $Y=-1$, dubbed the Higgs Triplet Model (HTM), offers an economical way to explain the tiny neutrino masses through the type-II seesaw mechanism~\cite{Konetschny:1977bn,Magg:1980ut,Schechter:1980gr,Cheng:1980qt,Lazarides:1980nt,Mohapatra:1980yp}. On the other hand, following the idea of thermal leptogenesis~\cite{Fukugita:1986hr}, the out-of-equilibrium decays of the heavy triplets in the early Universe 
generate the lepton number asymmetry~\cite{Ma:1998dx,Hambye:2003ka,Hambye:2005tk,Hambye:2012fh},\footnote{In order to generate CP violation, at least two triplet scalars are needed. Alternatively, one can also introduce one triplet scalar and one additional heavy Majorana neutrino, which is able to accommodate both the neutrino mass spectrum and the observed baryon asymmetry~\cite{Gu:2006wj}. Recently, it was pointed out that the inclusion of only one triplet scalar could fulfill successful leptogenesis through the Affleck-Dine mechanism~\cite{Affleck:1984fy} while the triplet could also play a role in inflation~\cite{Barrie:2021mwi,Barrie:2022cub}.} which can partly be converted to the baryon number asymmetry via the sphaleron process~\cite{Dashen:1974ck,Manton:1983nd,Forgacs:1983yu,Burzlaff:1983jb,Klinkhamer:1984di,Yaffe:1989ms}. In addition, the triplet scalar modifies the scalar potential of the SM and thus may change the pattern of the EW phase transition. Recently, it was found that there exists viable parameter space for a strong first-order EW phase transition in the HTM, and the spectrum of the produced gravitational waves was calculated~\cite{Zhou:2022mlz}. Nevertheless, it is still unclear whether or not a successful EW baryogenesis could be fulfilled in the framework of the HTM. To achieve this goal, a necessary step is to calculate the sphaleron configuration in the presence of a triplet scalar, which is the main purpose of the present work.

The sphaleron process plays a crucial role in dynamically generating the cosmological matter-antimatter asymmetry~\cite{Kuzmin:1985mm}. It is well known that the vacuum structure of non-Abelian gauge theories is nontrivial and the topologically distinct vacua are characterized by the Chern-Simons numbers~\cite{tHooft:1976rip,Callan:1976je,Jackiw:1976pf}, which can be directly related to the baryon ($B$) and lepton ($L$) numbers. Due to the chiral anomaly~\cite{Adler:1969gk,Bell:1969ts}, $B$ and $L$ are not conserved in the SM. The transition between two topologically distinct vacua changes the Chern-Simons number and hence $B$ and $L$ (but with $B-L$ conserved). The energy barrier between different vacua is characterized by the sphaleron energy $E^{}_{\rm sph}$. At zero temperature, we have $E^{}_{\rm sph}\sim 4\pi v/g \sim 5~{\rm TeV}$, where $v\approx 246~{\rm GeV}$ is the EW vacuum expectation value (VEV) and $g\approx 0.65$ is the ${\rm SU(2)}_{\rm L}^{}$ gauge coupling. Therefore, the $B$-violating sphaleron rate is highly suppressed at low temperatures: $\Gamma^{}_{\rm sph}\sim {\rm exp}\left(-E^{}_{\rm sph}/T\right)$~\cite{Arnold:1987mh,Arnold:1987zg,Carson:1990jm}. At temperatures above the EW scale, the VEV becomes zero and the energy barrier vanishes. In this case, the $B$-violating rate is no longer suppressed\footnote{Strictly speaking, there is no classical sphaleron solution above the critical temperature $T^{}_c$ of the EW phase transition. This is because the temperature-dependent VEV $v(T)$ turns out to be zero at $T>T^{}_c$ and the classical configuration scale $1/v(T)$ goes infinity. However, the $B$-violating process is still significant above $T^{}_c$ and the temperature provides a typical scale $\left(\alpha^{}_{\rm W} T\right)^{-1}$ for the sphaleron-like configuration~\cite{Arnold:1987mh,Arnold:1987zg}.} and is given by $\Gamma^{}_{\rm sph} \sim \alpha_{\rm W}^5 T^4$ with $\alpha^{}_{\rm W} \equiv g^2/\left(4\pi\right)$~\cite{Arnold:1996dy}. On the other hand, from the view of the classical field theory, the sphaleron configuration is the saddle-point solution of the energy functional~\cite{Dashen:1974ck,Manton:1983nd,Forgacs:1983yu,Burzlaff:1983jb,Klinkhamer:1984di,Yaffe:1989ms}. The sphaleron energy in the SM is mainly contributed by the Higgs and the gauge bosons. However, in the HTM, the triplet scalar has additional couplings to the gauge fields and to the SM Higgs field, hence is expected to influence the vacuum structure and the sphaleron configuration. As has been discussed above, the sphaleron energy plays an important role in both EW baryogenesis and leptogenesis. Therefore, it is necessary to recalculate the sphaleron configuration in the presence of a triplet scalar in order to realize a self-consistent baryogenesis in the framework of the HTM. 

The remaining part of this paper is organized as follows. In Sec.~\ref{sec:formalism}, we briefly review the minimax procedure to find the sphaleron solution and set up our formalism. In Sec.~\ref{sec:minmodel} and Sec.~\ref{sec:fullpotential}, we calculate the sphaleron configuration in the HTM, where a minimal version of the potential and a full potential is adopted, respectively. Our main conclusion is summarized in Sec.~\ref{sec:summary}, together with some further discussions. Finally, the numerical techniques to solve the equations of motion (EOM) of the sphaleron are provided in appendices.

\section{Theoretical Setup and Sphaleron Ansatz}
\label{sec:formalism}
In this section, we set up the general formalism to calculate the sphaleron configuration in the SM extended by a complex triplet scalar. We make the following two reasonable assumptions:
\begin{itemize}
\item The contribution from fermion fields to the sphaleron is neglected.
\item The finite Weinberg angle has little influence on the sphaleron (e.g., less than 1\% correction to the sphaleron energy)~\cite{Klinkhamer:1990fi,Kleihaus:1991ks,Kunz:1992uh,James:1992re}. Therefore, we can safely neglect the mixing between ${\rm SU}(2)_{\rm L}$ and ${\rm U}(1)_{\rm Y}$ gauge bosons such that the sphaleron configuration is spherically symmetric.
\end{itemize}

Under the above assumptions, the Lagrangian in the HTM is given by
\begin{eqnarray}
	\label{eq:lag}
	{\cal L}_{\rm HTM}=-\frac{1}{2}\tr\left(F^{}_{\mu \nu}F^{\mu\nu}_{}\right)+\left(D^{}_\mu \phi\right)^\dagger_{} \left(D^\mu_{} \phi\right)+\frac{1}{2}\tr\left[\left(D^\mu_{} \Delta\right)^\dagger_{}\left(D^{}_\mu \Delta\right)\right]-V(\phi,\Delta)\;.
\end{eqnarray}
The field strength in Eq.~(\ref{eq:lag}) is defined as $F^{}_{\mu\nu}=\partial^{}_\mu W^{}_\nu-\partial^{}_\nu W^{}_\mu-{\rm i}g\left[W^{}_\mu,W^{}_\nu\right]$, where $W^{}_\mu \equiv W_\mu^a \sigma^a_{}/2$ with $W_\mu^a$ the ${\rm SU}(2)^{}_{\rm L}$ gauge fields and $\sigma^a_{}$ (for $a=1,2,3$) the Pauli matrices. In addition, $D^{}_\mu$ is the covariant derivative, $\phi$ is the SM Higgs doublet, and $\Delta$ is the triplet scalar with hypercharge $Y=-1$ and transforms according to the adjoint representation of the ${\rm SU}(2)^{}_{\rm L}$ group
\begin{eqnarray}
\phi=\left(
\begin{matrix}
\phi^{+}_{}\\
\phi^0_{}
\end{matrix}
\right)\;,\qquad
\Delta=\left(
\begin{matrix}
\Delta^{-}_{}&-\sqrt{2}\Delta^0_{}\\
\sqrt{2}\Delta^{--}_{}&-\Delta^{-}_{}
\end{matrix}
\right)\;.
\end{eqnarray}
The VEVs of the scalar fields, namely $\langle \phi \rangle=v^{}_\phi/\sqrt{2}$ and $\langle \Delta \rangle= -v^{}_\Delta$, are determined by minimizing the scalar potential $V(\phi,\Delta)$, and satisfy $\sqrt{v_\phi^2+2v_\Delta^2}=v\approx 246~{\rm GeV}$. We will discuss it in more detail later. 

For the calculation of the sphaleron, since we are only focusing on the static field configuration, all the time components in Eq.~(\ref{eq:lag}) can consistently be set to zero. Then the energy density reads
\begin{eqnarray}
	\label{eq:energydensity}
	{\cal H}\left[W^{}_\mu,\phi,\Delta\right]=\frac{1}{2}g^{ik}_{}g^{jl}_{}{\rm Tr}\left(F^{}_{ij}F^{}_{kl}\right)+g^{ij}_{}\left(D^{}_i\phi\right)^\dagger_{}\left(D^{}_j\phi\right)+\frac{1}{2}g^{ij}_{}\left[\left(D^{}_i \Delta\right)^\dagger_{} \left(D^{}_j\Delta\right)\right]+V\left(\phi,\Delta\right)\;,\quad
\end{eqnarray}
where $g^{ij}_{}$ is the metric of the coordinate system. Since the sphaleron has a spherical symmetry in a pure ${\rm SU}(2)_{\rm L}$ gauge theory, it is most convenient to adopt the spherical coordinates $(r,\theta,\varphi)$. Then we have $g^{}_{ij}=\left(g^{ij}_{}\right)^{-1}_{}={\rm diag}\left(1,r^2_{},r^2_{}\sin^2_{}\theta\right)$. Moreover, the degrees of freedom from the gauge symmetry allow us to take the polar gauge. That is, the radial part of the gauge field can always be set to zero: $W^{}_r=0$. The total energy is determined by integrating over the whole space
\begin{eqnarray}
	\label{eq:energydef}
	E\left[W^{}_\mu,\phi,\Delta\right]=\int_{0}^{2\pi}{\rm d}\varphi\int_{0}^{\pi}{\rm d}\theta \sin\theta \int_0^{\infty}{\rm d}r\,r^2_{} \, {\cal H}\left[W^{}_\mu,\phi,\Delta\right] \;,
\end{eqnarray}
which is the functional of the field configuration.

Below we use the minimax procedure~\cite{Manton:1983nd,Manton:2004tk,Manton:2019qka} to find the sphaleron solution in the HTM. The basic idea is to construct a set of non-contractible loops\footnote{The loops are defined on the infinite-dimensional field configuration space $\left\{W_\mu({\bf x}),\phi({\bf x}),\Delta({\bf x})\right\}$, on which the energy functional $E\left[W_\mu({\bf x}),\phi({\bf x}),\Delta({\bf x})\right]$ is also defined. Here ${\bf x}$ denotes the general spatial indices.} starting and ending at the vacuum. For each of the loop there exists a configuration with maximum energy. Then the infimum of the maximum energies defines the sphaleron configuration, which corresponds to the saddle point of the energy functional. Along this line, the sphaleron configuration in the SM can be worked out~\cite{Manton:1983nd}. Similar strategies have also been used to study the sphaleron in the new-physics scenarios, which extend the SM by adding new singlet or doublet scalars~\cite{Kastening:1991ew,Choi:1994mf,Ahriche:2007jp,Fuyuto:2014yia,Enqvist:1992kd,Kastening:1991nw,Grant:2001at,Moreno:1996zm,Funakubo:2005bu}. However, as far as we know, the study of the sphaleron in the presence of a triplet scalar is still lacking. In what follows we show that the minimax procedure works in the HTM as well.

First, the fields at infinity ($r\to \infty$) should be related to the vacuum configuration via
\begin{eqnarray}
	W_j^\infty&=&-\frac{\rm i}{g}\partial^{}_j U^{}_\infty\left(\theta,\varphi\right)U_\infty^{-1}\left(\theta,\varphi\right)\;,\quad j=\theta,\varphi\;,\label{eq:gaugevacuum}\\
	\phi^\infty_{}&=&\frac{1}{\sqrt{2}}U_\infty\left(\theta,\varphi\right)^{}\left(
	\begin{matrix}
		0\\
		v^{}_\phi
	\end{matrix}
	\right)\;,\\
	\Delta^\infty_{}&=&U^{}_\infty\left(\theta,\varphi\right)\left(
	\begin{matrix}
		0&-v^{}_\Delta\\
		0&0
	\end{matrix}
	\right)U_\infty^{-1}\left(\theta,\varphi\right)\;,
	\label{eq:tripletvacuum}
\end{eqnarray}
where $U^{}_\infty\left(\theta,\varphi\right)\in {\rm SU}(2)^{}_{\rm L}$ denotes the gauge transformation that preserves the polar gauge condition. Note that Eq.~(\ref{eq:gaugevacuum}) satisfies the pure gauge such that the field strength $F_{\mu\nu}$ vanishes at the infinity, and Eq.~(\ref{eq:tripletvacuum}) comes from the fact that $\Delta$ belongs to the adjoint representation of ${\rm SU}(2)^{}_{\rm L}$. The gauge transformation $U_\infty(\theta,\varphi)$ (or equivalently, the Higgs field at infinity $\phi^\infty_{}$) defines a map: $S^2 \to S^3$ that is contractible, because the homotopy group $\pi^{}_2(S^3_{})$ is trivial. This implies that the fields at infinity can be continuously transformed to the vacuum configuration. In order to find a non-contractible loop in the field configuration space, we could introduce a new parameter $\mu\in [0,\pi]$, and extend the gauge transformation to 
\begin{eqnarray}
	\label{eq:gaugetrans}
	U\left(\mu,\theta,\varphi\right)=\left(
	\begin{matrix}
		e^{{\rm i}\mu}_{}\left(\cos\mu-{\rm i}\sin\mu \cos\theta\right)& e^{{\rm i}\varphi}_{}\sin\mu\sin\theta\\
		-e^{-{\rm i}\varphi}_{}\sin\mu\sin\theta&e^{-{\rm i}\mu}_{}\left(\cos\mu+{\rm i}\sin\mu\cos\theta\right)
	\end{matrix}
	\right)\;,
\end{eqnarray}
which satisfies $U\left(\mu,\theta=0,\varphi\right)=U\left(\mu=0,\theta,\varphi\right)=U\left(\mu=\pi,\theta,\varphi\right)=\mathbf{1}$ with $\mathbf{1}$ the identity matrix. Therefore, $\mu=0$ and $\mu=\pi$ correspond to the vacuum configuration, and the varying $\mu\in [0,\pi]$ parametrizes the loop. Then it follows that equipped with the loop parametrized by $\mu$, the gauge transformation $U(\mu,\theta,\varphi)$ defines a map: $S^3 \to S^3$. Since the homotopy group is $\pi_3(S^3)=\mathbb{Z}$, the topological degree of the map is nonzero and the loop is non-contractible. Now it is straightforward to construct the general field configuration using Eq.~(\ref{eq:gaugetrans}). A suitable ansatz is 
\begin{eqnarray}
	\label{eq:W}
	W^{}_j\left(\mu,r,\theta,\varphi\right)&=&-\frac{{\rm i}}{g}f(r)\partial^{}_j U\left(\mu,\theta,\varphi\right)U^{-1}_{}\left(\mu,\theta,\varphi\right)\;,\quad j=\theta,\varphi\;,\\
	\label{eq:phi}
	\phi\left(\mu,r,\theta,\varphi\right)&=&\frac{v^{}_\phi}{\sqrt{2}}h(r)U\left(\mu,\theta,\varphi\right)\left(
	\begin{matrix}
		0\\
		1
	\end{matrix}
	\right)\;,\\
	\label{eq:Delta}
	\Delta\left(\mu,r,\theta,\varphi\right)&=&v^{}_\Delta h^{}_\Delta(r) U\left(\mu,\theta,\varphi\right)\left(
	\begin{matrix}
		0&-1\\
		0&0
	\end{matrix}
	\right)U^{-1}\left(\mu,\theta,\varphi\right)\;,
\end{eqnarray}
where $f(r)$, $h(r)$ and $h^{}_\Delta(r)$ are radial profile functions to be determined. Since the polar gauge is singular at the origin, the smoothness requires the profile functions of all gauge multiplets to vanish at the origin. In addition, at spatial infinity the field configuration should go back to the vacuum configuration. This ensures the finiteness of the energy. Therefore, the boundary conditions of the profile functions should be
\begin{eqnarray}
	\label{eq:BC}
	f(0)&=&h(0)=h_\Delta(0)=0\;,\nonumber\\
	f(\infty)&=&h(\infty)=h^{}_\Delta(\infty)=1\;.
\end{eqnarray}

Substituting Eqs.~(\ref{eq:W})-(\ref{eq:Delta}) into Eq.~(\ref{eq:energydensity}), we obtain the kinematic terms
\begin{eqnarray}
\frac{1}{2}g^{ik}_{}g^{jl}_{}\tr\left(F^{}_{ij}F^{}_{kl}\right)&=&\frac{4}{g^2_{} r^4_{}}\sin^2_{}\mu\left[2f^2_{}\left(1-f\right)^2\sin^2_{}\mu+r^2_{} f'^2_{}\right]\;,\label{eq:kin1}\\
g^{ij}_{}\left(D^{}_i\phi\right)^\dagger_{}\left(D^{}_j\phi\right) &=&\frac{v_\phi^2}{2r^2_{}}\left[2\left(1-f\right)^2_{} h^2_{} \sin^2_{}\mu+r^2_{} h'^2_{}\right]\;,\label{eq:kin2}\\
\frac{1}{2}g^{ij}_{}\left[\left(D^{}_i \Delta\right)^\dagger_{} \left(D^{}_j\Delta\right)\right] &=& \frac{v_\Delta^2}{2r^2_{}}\left[\left(5-\cos2\theta\right)\left(1-f\right)^2_{}h_\Delta^2 \sin^2_{}\mu+r^2_{} h_\Delta'^2\right]\;,\label{eq:kin3}
\end{eqnarray}
where we have suppressed all arguments in the profile functions for simplicity, and all derivatives are with respect to $r$. It is interesting to notice that the kinetic terms of gauge fields and the doublet are spherically symmetric while that of the triplet is not. Also note that the contribution from the kinetic term of the triplet is suppressed by $v_\Delta^2/v_\phi^2$ compared with that of the doublet. Furthermore, once the scalar potential $V\left(\phi,\Delta\right)$ is known (as shown in the next two sections), one could obtain the total energy $E(\mu)$ by performing the integral in Eq.~(\ref{eq:energydef}), which is the function of the loop parameter $\mu$. The sphaleron configuration (labeled by $\mu^{}_0$) is determined by finding the maximum energy along the non-contractible loop, namely
\begin{eqnarray}
\frac{\delta E(\mu)}{\delta \mu}\Big|_{\mu=\mu_0} = 0\;,\qquad
\frac{\delta^2 E(\mu)}{\delta \mu^2}\Big|_{\mu=\mu_0} < 0\;.
\end{eqnarray} 
The sphaleron energy is given by $E_{\rm sph}=E(\mu_0)$, and the
EOM of the sphaleron are obtained from
\begin{eqnarray}
\frac{\delta E(\mu^{}_0)}{\delta f}=\frac{\delta E(\mu^{}_0)}{\delta h}=\frac{\delta E(\mu^{}_0)}{\delta h^{}_\Delta}=0\;.
\end{eqnarray}
Solving the EOM together with the boundary conditions in Eq.~(\ref{eq:BC}), one obtains the field configuration of the sphaleron. In the next two sections, we will use the above formalism to calculate the sphaleron configuration in the HTM.

\section{Sphaleron with the Minimal Potential}
\label{sec:minmodel}
\subsection{Scalar Potential}
The most general scalar potential in the HTM has 8 independent parameters. Before investigating the full potential in the next section, we first consider a simplified potential
\begin{eqnarray}
	\label{eq:minimalpotential}
	V\left(\phi,\Delta\right)=\lambda\left(\phi^\dagger_{} \phi\right)^2_{}-\kappa^2_{} \phi^\dagger_{} \phi+\frac{1}{2}M_\Delta^2 \tr\left(\Delta^\dagger_{} \Delta\right)-\left(\lambda^{}_\Delta M^{}_\Delta \phi^{\rm T}_{} \epsilon \Delta \phi+{\rm h.c.}\right)\;,
\end{eqnarray}
where $\epsilon\equiv {\rm i}\sigma^2_{}$. In Eq.~(\ref{eq:minimalpotential}), only the trilinear interaction ($\phi$-$\Delta$-$\phi$) is kept and all the quartic terms of triplet self-interaction and doublet-triplet interaction are turned off. This is a minimal version of the HTM, which still violates the lepton number and can accommodate the tiny neutrino masses. We will restrict ourselves to the minimal HTM throughout this section. It helps to exhibit the effects of the triplet on the sphaleron in a more apparent way.

Without loss of any generality, we can take $M_\Delta$ and $\lambda_\Delta$ in Eq.~(\ref{eq:minimalpotential}) to be real and positive. Substituting the VEVs into the scalar potential we have
\begin{eqnarray}
	\label{eq:minimalpotential2}
	V\left(v^{}_\phi,v^{}_\Delta\right)\equiv V\left(\langle\phi\rangle,\langle\Delta\rangle\right)=\frac{1}{4}\lambda v_\phi^4-\frac{1}{2}\kappa^2_{} v_\phi^2+\frac{1}{2}M_\Delta^2 v_\Delta^2-\lambda^{}_\Delta M^{}_\Delta v^{}_\Delta v_\phi^2\;.
\end{eqnarray}
The VEVs are determined by minimizing the potential
\begin{eqnarray}
	&&\frac{\partial}{\partial v^{}_\phi}V\left(v^{}_\phi,v^{}_\Delta\right)=\lambda v_\phi^3-\kappa^2_{} v^{}_\phi-2\lambda^{}_\Delta M^{}_\Delta v^{}_\Delta v^{}_\phi=0\;,\\
	&&\frac{\partial}{\partial v^{}_\Delta}V\left(v^{}_\phi,v^{}_\Delta\right)=M_\Delta^2 v^{}_\Delta-\lambda^{}_\Delta M^{}_\Delta v_\phi^2=0\;,
\end{eqnarray}
from which one obtains
\begin{eqnarray}
	\label{eq:vev}
	v^{}_\phi=\sqrt{\frac{\kappa^2_{}}{\lambda-2\lambda_\Delta^2}}\;,\quad
	v_\Delta^{}=\frac{\lambda^{}_\Delta v_\phi^2}{M^{}_\Delta}\;.
\end{eqnarray}
In order to have a real positive $v_\phi$, we require $\kappa^2_{}>0 $ and $\lambda-2\lambda_\Delta^2>0$. Besides, the vacuum stability requires $\lambda>0$. Substituting the VEVs back to Eq.~(\ref{eq:minimalpotential2}) we obtain the minimum 
\begin{eqnarray}
	V^{}_{\rm min}=-\frac{\kappa^4_{}}{4\left(\lambda-2\lambda_\Delta^2\right)}=-\frac{1}{4}\left(\lambda-2\lambda_\Delta^2\right) v_\phi^4\;.
\end{eqnarray}
The nonzero minimum of the potential would bring about infinity after integrating over the whole space. To obtain a finite energy, one can perform a constant shift to the potential
\begin{eqnarray}
	\label{eq:minimalpotential_shift}
	V\left(\phi,\Delta\right)&\to& V\left(\phi,\Delta\right)+\frac{1}{4}\left(\lambda-2\lambda_\Delta^2\right) v_\phi^4\nonumber\\
	&=&\lambda\left(\phi^\dagger_{}\phi-\frac{v_\phi^2}{2}\right)^2_{}+2\lambda_\Delta^2 v_\phi^2\left(\phi^\dagger_{}\phi-\frac{v_\phi^2}{2}\right)+\frac{\lambda_\Delta^2 v_\phi^4}{2v_\Delta^2}\left[\tr\left(\Delta^\dagger_{} \Delta\right)-v_\Delta^2\right]\nonumber\\
	&&+\frac{\lambda_\Delta^2 v_\phi^2}{v^{}_\Delta}\left[v^{}_\Delta v_\phi^2-2\,{\rm Re} \left(\phi^{\rm T}_{} \epsilon \Delta \phi\right)\right]\;.
\end{eqnarray}
Note that such a shift has no impact on the sphaleron configuration since it does not involve any dynamical degrees of freedom. In Eq.~(\ref{eq:minimalpotential_shift}) we have replaced $\kappa^2_{}$ and $M^{}_\Delta$ with the VEVs using Eq.~(\ref{eq:vev}). Therefore, in the minimal HTM the scalar potential depends on 4 real positive parameters: $\left\{\lambda,\lambda^{}_\Delta,v^{}_\phi,v^{}_\Delta\right\}$. Substituting Eqs.~(\ref{eq:W})-(\ref{eq:Delta}) into Eq.~(\ref{eq:minimalpotential_shift}), we get the scalar potential in terms of the profile functions
\begin{eqnarray}
	\label{eq:potential}
V\left(\phi,\Delta\right)=\frac{1}{4}v_\phi^4\left[\lambda\left(1-h^2_{}\right)^2_{}+2\lambda_\Delta^2\left(2h^2_{}-1-h^{}_\Delta\right)\left(1-h^{}_\Delta\right)\right]\;.
\end{eqnarray}
It can be seen that the scalar potential is also spherically symmetric, although the fields themselves (i.e., $\phi$ and $\Delta$) are not.

\subsection{Equations of Motion}
Now one can calculate the total energy using Eq.~(\ref{eq:energydef}). It is helpful to define the following dimensionless quantity
\begin{eqnarray}
	\label{eq:xidef}
\xi \equiv g v r \approx 8.1 \times \left(\frac{r}{10^{-15}~{\rm cm}}\right)\;,
\end{eqnarray}
where we have used $g\approx 0.65$ and $v=\sqrt{v_\phi^2+2v_\Delta^2}\approx 246~{\rm GeV}$.
As one can see later, $\xi$ characterizes the typical scale of the sphaleron. Substituting Eqs.~(\ref{eq:kin1})-(\ref{eq:kin3}) and (\ref{eq:potential}) into Eq.~(\ref{eq:energydef}) and integrating out the angular part, we obtain
\begin{eqnarray}
	\label{eq:energymu}
	E(\mu)=\frac{4\pi v}{g}\int_{0}^{\infty}{\rm d}\xi \left({\cal H}^{}_{\rm gauge}+{\cal H}^{}_{\rm doublet}+{\cal H}^{}_{\rm triplet}\right)\;,
\end{eqnarray}
where\footnote{From here on, unless otherwise specified, all derivatives are with respect to $\xi$.}
\begin{eqnarray}
	\label{eq:E1}
	{\cal H}^{}_{\rm gauge}&=&4 f'^2_{}\sin^2_{}\mu+\frac{8}{\xi^2_{}}f^2_{}\left(1-f\right)^2\sin^4_{}\mu\;,\\
	\label{eq:E2}
	{\cal H}^{}_{\rm doublet}&=&\frac{\varrho^{}_{1}}{4\beta^2_{}}\xi^2_{}\left(1-h^2_{}\right)^2+\frac{1}{2\beta}\xi^2_{} h'^2_{}+\frac{1}{\beta}h^2_{}\left(1-f\right)^2\sin^2_{}\mu\;,\\
	\label{eq:E3}
	{\cal H}^{}_{\rm triplet}&=&\frac{\varrho_{2}}{4\beta^2_{}}\xi^2\left(2h^2_{}-1-h^{}_\Delta\right)\left(1-h^{}_\Delta\right)+\frac{\varrho^{}_{3}}{6\beta}\left[3\xi^2_{} h_\Delta'^2+16h_\Delta^2\left(1-f\right)^2\sin^2_{}\mu\right]\;,
\end{eqnarray}
and
\begin{eqnarray}   
	\varrho^{}_{1} \equiv \frac{\lambda}{g^2_{}}\;,\quad
	\varrho^{}_{2} \equiv \frac{2\lambda_\Delta^2}{g^2_{}}\;,\quad
	\varrho^{}_{3} \equiv \frac{v_\Delta^2}{v_\phi^2}\;,\quad
	\beta\equiv \frac{v^2_{}}{v_\phi^2}=1+2\varrho^{}_{3}\;.
\end{eqnarray}
In Eq.~(\ref{eq:energymu}) we have divided the contributions into three parts: ${\cal H}^{}_{\rm gauge}$ and ${\cal H}^{}_{\rm doublet}$ come from the kinetic and self-interaction terms of the gauge bosons and the doublet, respectively, while ${\cal H}^{}_{\rm triplet}$ arises from the triplet kinetic term, the triplet mass term, and the doublet-triplet interaction. To reduce to the SM case, one can simply take $\varrho^{}_{2}=\varrho^{}_{3}=0$. 

The next step is to determine the value of $\mu$ corresponding to the maximum energy. To this end, we calculate the variation of the energy with respective to $\mu$, i.e.,
\begin{align}
	\frac{\delta E(\mu)}{\delta \mu}=\frac{4\pi v}{3g}\sin2\mu\int_{0}^{\infty}{\rm d}\xi\left[12f'^2_{}+\frac{1}{\beta}\left(1-f\right)^2_{}\left(3h^2_{}+8\varrho^{}_{3} h_\Delta^2\right)+\frac{48}{\xi^2_{}}f^2_{}\left(1-f\right)^2_{}\sin^2_{}\mu\right]=0\;,
\end{align}
which gives $\mu=0$, $\pi/2$ or $\pi$. A further investigation of the second-order variation leads to
\begin{align}
	\frac{\delta^2_{} E(\mu)}{\delta \mu^2}\bigg|^{}_{\mu=0}&=\frac{\delta^2_{} E(\mu)}{\delta \mu^2_{}}\bigg|^{}_{\mu=\pi}=\frac{4\pi v}{g}\int_{0}^{\infty}{\rm d}\xi\left[8f'^2_{}+\frac{2}{3\beta}\left(1-f\right)^2\left(3h^2_{}+8\varrho^{}_{3} h_\Delta^2\right)\right]>0\;,\\
	\frac{\delta^2_{} E(\mu)}{\delta \mu^2_{}}\bigg|^{}_{\mu=\pi/2}&=\frac{4\pi v}{g}\int_{0}^{\infty}{\rm d}\xi\left[-8f'^2_{}-\frac{2}{3\beta}\left(1-f\right)^2_{}\left(3h^2_{}+8\varrho^{}_{3} h_\Delta^2\right)-\frac{32}{\xi^2_{}}f^2_{}\left(1-f\right)^2_{}\right]<0\;.
\end{align}

Therefore, $\mu=0$ or $\pi$ corresponds to the minimum energy (i.e., the vacuum configuration) as expected, while $\mu=\pi/2$ corresponds to the maximum energy (i.e., the sphaleron configuration). Substituting $\mu=\pi/2$ into Eq.~(\ref{eq:energymu}) we obtain the sphaleron energy
\begin{align}
	E^{}_{\rm sph}=\frac{4\pi v}{g}\int_{0}^{\infty}{\rm d}\xi&\left\{
	4f'^2_{}+\frac{8}{\xi^2_{}}f^2_{}\left(1-f\right)^2_{}+\frac{1}{\beta}\left(1-f\right)^2_{}h^2_{}+\frac{1}{2\beta}\xi^2_{} h'^2_{}+\frac{\varrho^{}_{3}}{6\beta}\left[3\xi^2_{} h_\Delta'^2+16h_\Delta^2\left(1-f\right)^2_{}\right]\right.\nonumber\\
	&\left.+\frac{\xi^2_{}}{4\beta^2_{}}\left[\left(\varrho^{}_{1}-\varrho^{}_{2}\right)\left(1-h^2_{}\right)^2_{}+\varrho^{}_{2}\left(h^2_{}-h^{}_\Delta\right)^2_{}\right]
	\right\}\;.
	\label{eq:energysph}
\end{align}
The EOM of the fields are determined by the variation of the sphaleron energy with respect to the profile functions 
\begin{eqnarray}
	\label{eq:EOMgeneral}
\frac{\delta E^{}_{\rm sph}}{\delta f}=\frac{\delta E^{}_{\rm sph}}{\delta h}=\frac{\delta E^{}_{\rm sph}}{\delta h^{}_\Delta}=0\;,
\end{eqnarray}
which results in 
\begin{eqnarray}
	\label{eq:EOM1}
	\xi^2_{} f''&=& 2 f\left(1-f\right)\left(1-2f\right)-\frac{\xi^2_{}}{4\beta}\left(1-f\right)h^2_{}-\frac{2\varrho^{}_{3}}{3\beta}\xi^2_{}\left(1-f\right)h_\Delta^2\;,\\
	\label{eq:EOM2}
	\left(\xi^2_{} h'\right)'&=&2\left(1-f\right)^2_{} h-\frac{\xi^2_{}}{\beta}\left[\left(\varrho^{}_{1}-\varrho^{}_{2}\right)h\left(1-h^2_{}\right)-\varrho^{}_{2}h\left(h^2_{}-h^{}_\Delta\right)\right]\;,\\
	\label{eq:EOM3}
	\left(\xi^2_{} h_\Delta'\right)'&=&\frac{16}{3}\left(1-f\right)^2_{}h^{}_\Delta-\frac{\varrho^{}_{2}}{2\beta\varrho^{}_{3}}\xi^2_{}\left(h^2_{}-h^{}_\Delta\right)\;.
\end{eqnarray}
In addition, the profile functions should satisfy the boundary conditions in Eq.~(\ref{eq:BC}). Once the solutions of the EOM are found, one can simply substitute them back to Eq.~(\ref{eq:energysph}) to get the sphaleron energy, which is expected to be of the order of $4\pi v/g\approx 5~{\rm TeV}$.

Before solving Eqs.~(\ref{eq:EOM1})-(\ref{eq:EOM3}), it is interesting to first take a look at the heavy-mass limit of the triplet scalar (i.e., $M_\Delta\to\infty$ or $v^{}_\Delta/v^{}_\phi\to 0$). Note that the coupling $\varrho^{}_2/(2\varrho^{}_{3})$ in Eq.~(\ref{eq:EOM3}) is actually $M_\Delta^2/(g^2 v_\phi^2)$ using the second relation in Eq.~(\ref{eq:vev}). In the heavy-mass limit, $M_\Delta^2/(g^2_{} v_\phi^2)$ goes infinity and Eq.~(\ref{eq:EOM3}) enforces $h^{}_\Delta \to h^2_{}$. Then the EOM of $f(\xi)$ and $h(\xi)$ reduce to
\begin{eqnarray}
	\xi^2_{} f''&=& 2f\left(1-f\right)\left(1-2f\right)-\frac{\xi^2_{}}{4}\left(1-f\right)h^2_{}\;,\\
	\left(\xi^2_{} h'\right)'&=&2\left(1-f\right)^2_{} h-\xi^2_{}\left(\varrho^{}_{1}-\varrho^{}_{2}\right)h\left(1-h^2_{}\right)\;,
\end{eqnarray}
which are exactly those in the SM~\cite{Klinkhamer:1984di}, except for the replacement $\varrho^{}_{1} \to \varrho^{}_{1}-\varrho^{}_{2}$, or equivalently, $\lambda\to \lambda_{\rm eff}\equiv \lambda-2\lambda_\Delta^2$. Therefore, a very heavy triplet scalar has no influence on the sphaleron but only shifts the quartic Higgs coupling $\lambda$ to $\lambda^{}_{\rm eff}$. This is consistent with the result that one integrates out the triplet scalar at the tree level and retains only the leading-order term:
\begin{eqnarray}
	{\cal L}^{}_{\rm eff}={\cal L}^{}_{\rm SM}+2\lambda_\Delta^2 \left(\phi^\dagger_{} \phi\right)^2_{}+{\cal O}\left(\frac{1}{M^{}_\Delta}\right)\;.
\end{eqnarray}
The study of the sphaleron configuration in the framework of effective field theories have been carried out in Refs.~\cite{Spannowsky:2016ile,Gan:2017mcv}.

\subsection{Sphaleron Solution}
\begin{figure*}[t!]
	\centering
	\includegraphics[scale=0.5]{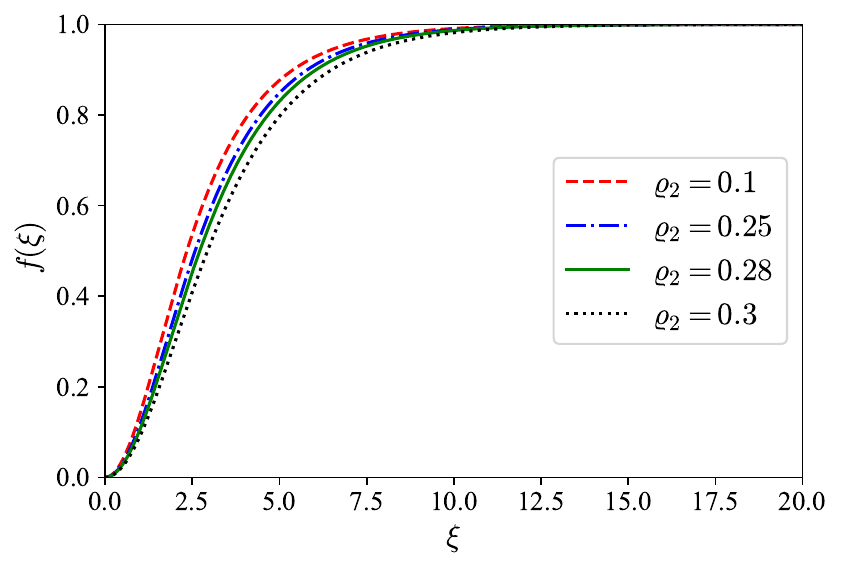}\quad
	\includegraphics[scale=0.5]{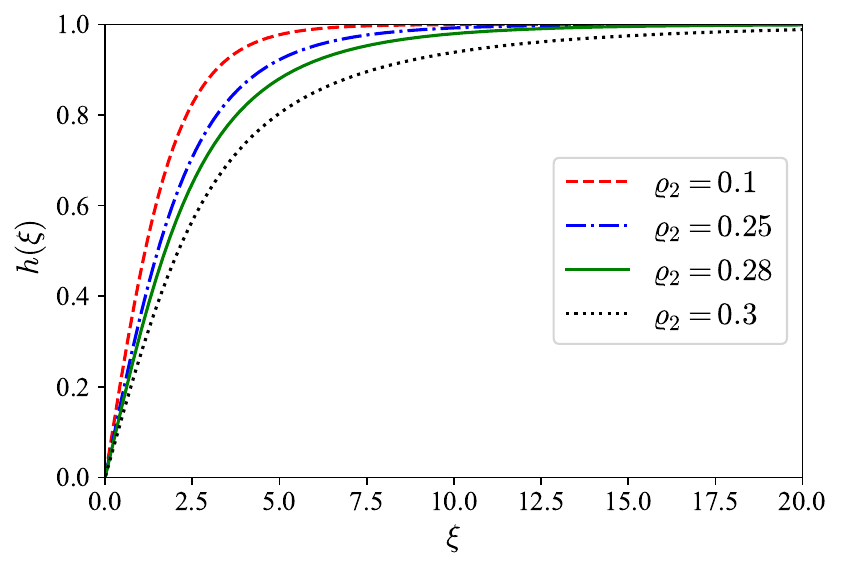}\quad\\
	\vspace{0.2cm}
	\includegraphics[scale=0.5]{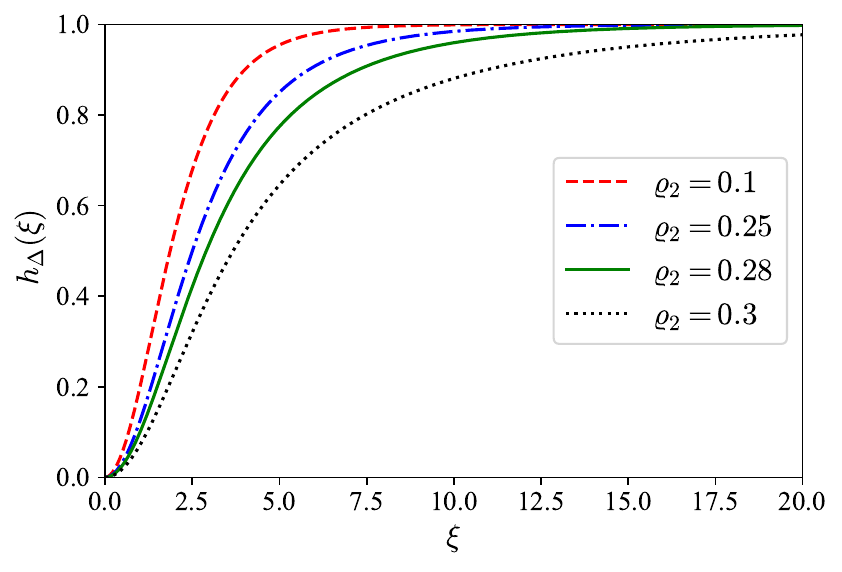}\quad
	\includegraphics[scale=0.5]{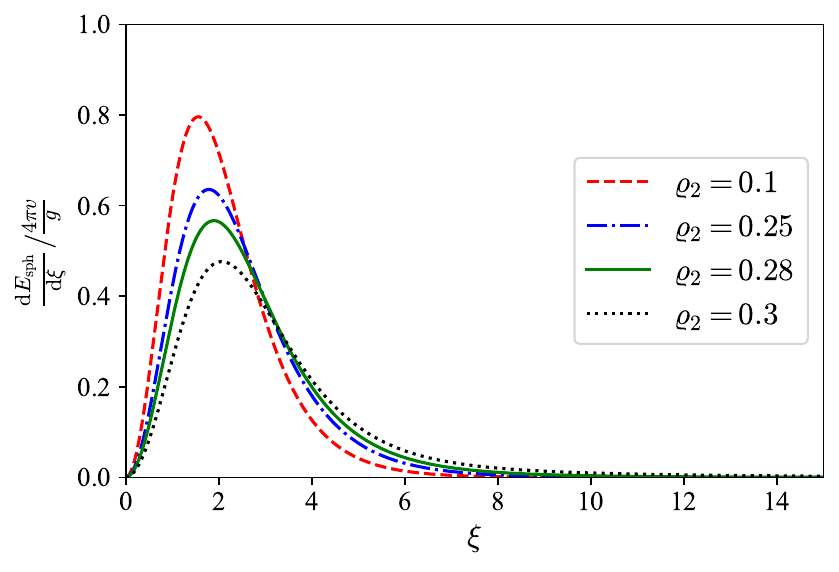}
	\caption{\label{fig:profile-HTM-min}The profile functions and sphaleron energy density in the minimal HTM are shown for different values of the doublet-triplet trilinear coupling parameter $\varrho^{}_{2}$, where $\varrho^{}_3 = 10^{-3}_{}$ and  $\varrho^{}_1 = 0.306$ have been taken (see the main text for more information).} 
\end{figure*}

The EOM in Eqs.~(\ref{eq:EOM1})-(\ref{eq:EOM3}) are coupled nonlinear differential equations. It is difficult to solve them analytically. In Appendix~\ref{app:spectral}, we have developed a numerical algorithm based on the spectral method that can be used to efficiently solve the sphaleron EOM. See Appendix~\ref{app:spectral} for more details.

The solutions of the profile functions and the sphaleron energy density obtained from the spectral method are shown in Fig.~\ref{fig:profile-HTM-min}. Note that $\varrho_{3}$ violates the custodial symmetry and thus is strictly constrained by the EW precision measurements: $\sqrt{\varrho^{}_{3}} = v^{}_\Delta/v^{}_\phi \lesssim 0.03$~\cite{ParticleDataGroup:2022pth}. Moreover, in the SM, $\varrho^{}_{1}$ is related to the mass ratio of the Higgs boson and $W$ boson via $\varrho_{1}^{\rm SM}=m_h^2/\left(8m_{\rm W}^2\right)\approx 0.306$. In Fig.~\ref{fig:profile-HTM-min}, as an illustration, we have taken $\varrho^{}_{3}$ to saturate the experimental upper bound, namely $\varrho^{}_{3}=10^{-3}_{}$ (corresponding to $v^{}_\Delta \approx 8~{\rm GeV}$). We also fix $\varrho^{}_{1}=\varrho_{1}^{\rm SM}$ and show the solutions of profile functions and the sphaleron energy density for different $\varrho^{}_{2}$. 

From Fig.~\ref{fig:profile-HTM-min}, it can be seen that all the profile functions approach the vacuum configuration [i.e., $f(\infty)=h(\infty)=h^{}_\Delta(\infty)=1$] quickly. The sphaleron energy is restricted within a very narrow region: $\xi \lesssim 10$, corresponding to $r\lesssim 10^{-15}_{}~{\rm cm}$ using Eq.~(\ref{eq:xidef}), which is even two orders of magnitude smaller than the length scale of a proton. This implies that the sphaleron looks like a ``particle" localized near the origin. If the triplet couples with the doublet, then a larger trilinear coupling  $\varrho^{}_{2}$ makes the profile functions tend to the vacuum configuration more slowly. In addition, $\varrho^{}_{2}$ would diffuse the distribution of the sphaleron energy density and also decrease the total energy of the sphaleron. 

It is also interesting to investigate the asymptotic behavior of the triplet field near the origin. First, from Eqs.~(\ref{eq:EOM1}) and (\ref{eq:EOM2}), the smoothness of the profile functions at the origin requires $f$ and $h$ to satisfy $f\sim \xi^2$ and $h\sim \xi$, which is the same as the SM case~\cite{Klinkhamer:1984di}. Then suppose $h^{}_\Delta \sim \xi^\alpha_{}$ (with $\alpha>0$) near $\xi=0$ and substitute it into Eq.~(\ref{eq:EOM3}). If $\varrho^{}_{3}\neq 0$, keeping only the leading-order term of $\xi$ one obtains\footnote{If $\varrho^{}_{3}=0$, the term proportional to $\xi^2_{}/\varrho^{}_{3}$ in Eq.~(\ref{eq:EOM3}) cannot be neglected near $\xi=0$. Instead, the finiteness of the both sides of Eq.~(\ref{eq:EOM3}) enforces $h_\Delta \to h^2$. Therefore we have $h_\Delta \sim h^2\sim \xi^2$ near the origin if $\varrho^{}_{3}=0$.} 
\begin{eqnarray}
	\alpha\left(\alpha-1\right)+2\alpha=\frac{16}{3}\;\;\Rightarrow\;\; \alpha=\frac{1}{6}\left(\sqrt{201}-3\right)\approx 1.86\;.
\end{eqnarray}
The above asymptotic behavior of the triplet field near the origin has also been verified numerically.

\begin{figure*}[t!]
	\centering
	\includegraphics[scale=0.6]{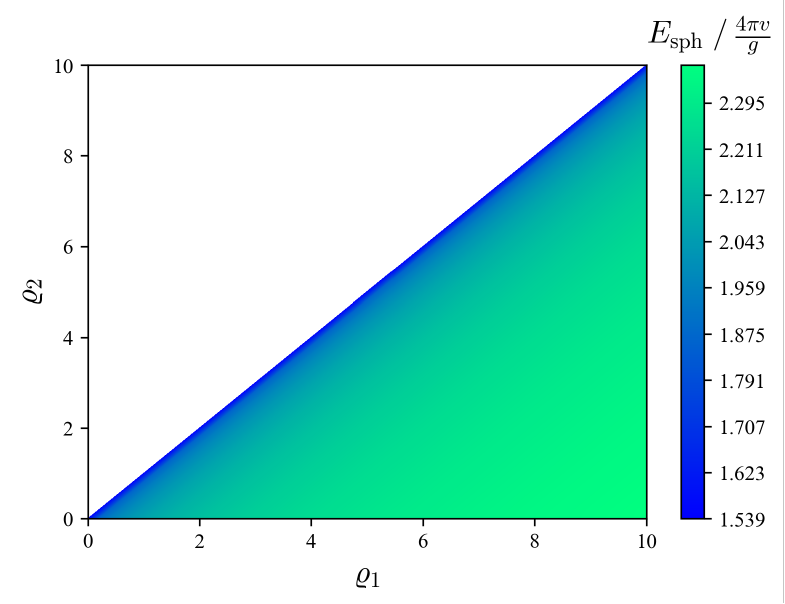}\quad
	\includegraphics[scale=0.6]{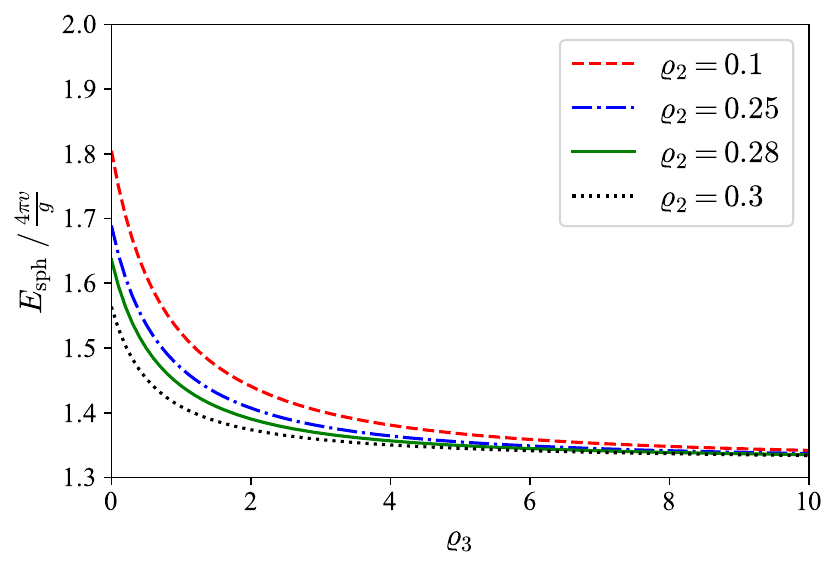}
	\caption{\label{fig:energy-HTM-min}The sphaleron energy in the minimal HTM versus coupling parameters. \emph{Left}: The contour plot of the sphaleron energy with respect to $\varrho^{}_{1}$ and $\varrho^{}_{2}$ with $\varrho^{}_{3}=10^{-3}$ being fixed. Note that $\varrho^{}_{1}\geqslant\varrho^{}_{2}$ is required from the EW vacuum stability. \emph{Right}: The sphaleron energy for different values of $\varrho^{}_{2}$ and $\varrho^{}_{3}$, where $\varrho^{}_{1}=0.306$ is taken.}
\end{figure*}

\begin{table}[t!]
	\normalsize 
	\setlength{\tabcolsep}{6pt}
	\renewcommand{\arraystretch}{1.2}
	\centering
	\begin{tabular}{l|c|l|c}
		\hline \hline
		$\varrho_{1}^{}$ &  $E_{\rm sph}^{\rm SM}$ & $\varrho_{1}-\varrho^{}_{2}$ & $E_{\rm sph}^{\rm HTM}$ \\
		\hline \hline
		0 &  1.5395 & 0  & 1.5391\\
		\hline
		0.001 &  1.5651 & 0.001 & 1.5646\\
		\hline
		0.01 &  1.6389 &0.01 & 1.6382\\
		\hline
		0.1 & 1.7994 &0.1 & 1.7985\\
		\hline
		0.2 & 1.8695 &0.2 & 1.8684\\
		\hline
		0.5 &  1.9766 &0.5 & 1.9754\\
		\hline
		1.0 &  2.0659 &1.0 & 2.0646\\
		\hline
		2.0 & 2.1589 & 2.0 & 2.1574\\
		\hline
		5.0 & 2.2800 & 5.0 & 2.2785\\
		\hline
		10.0 & 2.3647 & 10.0 & 2.3639\\
		\hline
		\hline
	\end{tabular}
	\caption{\label{table:compare}Comparison of the sphaleron energy between the SM and the minimal HTM (with $\varrho^{}_{3}=10^{-3}_{}$). 
	All numerical results are obtained using the spectral method developed in Appendix~\ref{app:spectral}. The first two columns denote the parameter and the corresponding sphaleron energy in the SM, while the last two columns denote those in the minimal HTM. Note that $E_{\rm sph}^{\rm HTM}$ only depends on the difference between $\varrho^{}_{1}$ and $\varrho^{}_{2}$ for such a small $\varrho^{}_{3}$. All energies are in units of $4\pi v/g$.}
\end{table}

In the left panel of Fig.~\ref{fig:energy-HTM-min}, we show the contour plot of the sphaleron energy with respect to $\varrho^{}_{1}$ and $\varrho^{}_{2}$, where $\varrho^{}_{3}=10^{-3}_{}$ is fixed. It is obvious that a larger $\varrho^{}_{1}$ (or $\varrho^{}_{2}$) would increase (or decrease) the sphaleron energy. One may wonder how large is the difference of the sphaleron energy between the minimal HTM and the SM. The answer is that for $\varrho^{}_{3}\lesssim 10^{-3}_{}$ the difference is negligible. This is because for such a small $\varrho^{}_{3}$, the triplet almost decouples and shifts $\lambda$ to $\lambda-2\lambda_\Delta^2$. As a result, the sphaleron energy in the minimal HTM only depends on $\varrho^{}_{1}-\varrho^{}_{2}$, as is shown in the left panel of Fig.~\ref{fig:energy-HTM-min}. In Table~\ref{table:compare}, we compare the sphaleron energy in the SM and in the minimal HTM. As one can see, the difference is only about 1\textperthousand, if one replaces $\varrho_{1}$ in the SM with $\varrho^{}_{1}-\varrho^{}_{2}$ in the minimal HTM. Note that such a difference is of the same order of $\varrho^{}_{3}$.

However, things are different for a larger $\varrho^{}_{3}$.\footnote{We comment here that a large value of $v^{}_\Delta/v^{}_\phi$ may be available when taking into account the temperature corrections in the early Universe. See more discussions in Sec.~\ref{sec:summary}.} In the right panel of Fig.~\ref{fig:energy-HTM-min} we show the behavior of $E^{}_{\rm sph}$ with $\varrho_{3}$. It can be seen that a large $\varrho^{}_{3}$ could significantly decrease the sphaleron energy. This can be understood as follows. For small $\varrho^{}_{3}$, $\beta \approx 1$, $h^{}_\Delta\approx h^2_{}$, and the term proportional to $\varrho^{}_{3}$ in Eq.~(\ref{eq:energysph}) is suppressed, which means the contribution of the triplet to the sphaleron energy is negligible, and it reduces to the SM case. However, for large $\varrho^{}_{3}$ we have $\beta\approx 2\varrho^{}_{3}$, then the terms relevant to the doublet in Eq.~(\ref{eq:energysph}) are suppressed by the inverse power of $\beta$. In this case, the sphaleron energy is dominated by the contribution of gauge fields and the triplet. More explicitly, we have
\begin{align}
	E^{}_{\rm sph} \left(\varrho^{}_{3}\gg 1\right)\approx \frac{4\pi v}{g}\int_{0}^{\infty}{\rm d}\xi\left\{
	4f'^2_{}+\frac{8}{\xi^2_{}}f^2_{}\left(1-f\right)^2_{}+\frac{1}{12}\left[3\xi^2_{} h_\Delta'^2+16h_\Delta^2\left(1-f\right)^2_{}\right]
	\right\}\approx 1.32\times \frac{4\pi v}{g}\;,
\end{align}
which tends to a fixed value. This explains why curves with different $\varrho^{}_{2}$ in the right panel of Fig.~\ref{fig:energy-HTM-min} converge together in the large $\varrho^{}_{3}$ limit. Compared with the case of small $\varrho^{}_{3}$, we find the sphaleron energy could be decreased by 30\% if $\varrho^{}_{3}$ is sufficiently large.

To summarize, in the minimal HTM, there are three relevant parameters which could affect the sphaleron configuration, i.e., the doublet quartic coupling $\varrho^{}_{1}$, the doublet-triplet trilinear coupling $\varrho_{2}$, and the VEV-ratio parameter $\varrho^{}_{3}$. As in the SM, the sphaleron energy increases monotonically with $\varrho^{}_{1}$, while the two additional parameters $\varrho^{}_{2}$ and $\varrho^{}_{3}$ would decrease the sphaleron energy. However, at zero temperature, the stringent constraint on the triplet VEV has highly suppressed the effects of the triplet on the sphaleron. The sphaleron energy in the minimal HTM can be simply obtained from that in the SM with the replacement $\varrho^{}_{1}\to \varrho^{}_{1}-\varrho^{}_{2}$.
As we will see below, the situation becomes different when considering the full potential in the HTM.

\section{Sphaleron with the Full Potential}
\label{sec:fullpotential}
In this section, we calculate the sphaleron configuration in the HTM with the full potential.

\subsection{Scalar Potential and Equations of Motion}
The most general scalar potential in the HTM is given by
\begin{align}
	\label{eq:fullpotential}
	V\left(\phi,\Delta\right)=&\lambda\left(\phi^\dagger_{} \phi\right)^2_{}-\kappa^2_{} \phi^\dagger_{} \phi+\frac{1}{2}M_\Delta^2 \tr\left(\Delta^\dagger_{} \Delta\right)-\left(\lambda^{}_\Delta M^{}_\Delta \phi^{\rm T}_{} \epsilon \Delta \phi+{\rm h.c.}\right)\notag\\
	&+\frac{\lambda_{1}}{4}\left[\tr\left(\Delta^{\dagger}_{}\Delta\right)\right]^{2}_{}+\frac{\lambda^{}_{2}}{4}\tr\left[\left(\Delta^{\dagger}_{}\Delta\right)^{2}_{}\right]+\lambda^{}_{3}\left(\phi^{\dagger}_{}\phi\right)\tr\left(\Delta^{\dagger}_{}\Delta\right)+\lambda^{}_{4}\phi^{\dagger}_{}\Delta\Delta^{\dagger}_{}\phi\;,
\end{align}
where $\lambda^{}_i$ (for $i=1,2,3,4$) are real couplings. Substituting the VEVs of the doublet and the triplet into the potential above and minimizing it leads to
\begin{align}
	\label{eq:vev1}
	\frac{\partial}{\partial v^{}_\phi}V\left(v^{}_\phi,v^{}_{\Delta}\right)&=\left(-\kappa^{2}_{}+\lambda v_\phi^{2}-2\lambda^{}_{\Delta}M^{}_{\Delta}v^{}_{\Delta}+\lambda^{}_{3}v_{\Delta}^{2}\right)v^{}_\phi=0\;, \\
	\label{eq:vev2}
	\frac{\partial}{\partial v^{}_{\Delta}}V\left(v^{}_\phi,v^{}_{\Delta}\right)&=-\lambda^{}_{\Delta}M^{}_{\Delta}v_\phi^{2}+M_{\Delta}^{2}v^{}_{\Delta}+(\lambda^{}_{1}+\lambda^{}_{2})v_{\Delta}^{3}+\lambda^{}_{3}v_\phi^{2}v^{}_{\Delta}=0\;.
\end{align}
From Eqs.~(\ref{eq:vev1}) and (\ref{eq:vev2}) one can determine $v^{}_\phi$ and $v^{}_\Delta$ from the couplings, though the general expressions are very tedious. Alternatively, we could also use Eqs.~(\ref{eq:vev1}) and (\ref{eq:vev2}) to express the couplings as
\begin{align}
	\label{eq:lambda3}
	\lambda^{}_{3}&=\frac{\kappa^{2}_{}-\lambda v_\phi^{2}+2\lambda^{}_{\Delta}M^{}_{\Delta}v^{}_{\Delta}}{v_{\Delta}^{2}}\;,\\
	\label{eq:lambda12}
	\lambda^{}_{1}+\lambda^{}_{2}&=-\frac{M^{}_\Delta}{v_\Delta^3}\left(v^{}_\Delta M^{}_\Delta + \lambda^{}_\Delta v_\phi^2 \right)+\frac{v_\phi^2}{v_\Delta^4}\left(\lambda v_\phi^2 - \kappa^2_{}\right)\;.
\end{align}
With the help of Eqs.~(\ref{eq:lambda3}) and (\ref{eq:lambda12}), the vacuum energy is given by
\begin{align}
	V\left(v_\phi,v_{\Delta}\right)=\frac{1}{4}\left[M^{}_{\Delta}v^{}_{\Delta}(M^{}_{\Delta}v^{}_{\Delta}-\lambda^{}_{\Delta}v_\phi^{2})-\kappa^{2}_{}v_\phi^{2}\right]\;.
\end{align}
As what we have done before, in order to have a finite total energy,
we perform a shift to the potential to make the vacuum energy being zero
\begin{align}
	V\left(\phi,\Delta\right) &\rightarrow V\left(\phi,\Delta\right)-\frac{1}{4}\left[M^{}_{\Delta}v^{}_{\Delta}\left(M^{}_{\Delta}v^{}_{\Delta}-\lambda^{}_{\Delta}v^{2}_{}\right)-\kappa^{2}_{}v_\phi^{2}\right]\nonumber\\
	=&+\lambda\left[\left(\phi^\dagger_{} \phi\right)-\frac{v_\phi^2}{2}\right]^2_{}+\left(\lambda v_\phi^2-\kappa^2_{}\right)\left[\left(\phi^\dagger_{} \phi\right)-\frac{v_\phi^2}{2}\right]+\frac{1}{2}M_\Delta^2 \left[\tr\left(\Delta^\dagger_{} \Delta\right)-v_\Delta^2\right]\nonumber\\
	&-\lambda^{}_\Delta M^{}_\Delta \left[2\,\Rea\left(\phi^{\rm T}_{} \epsilon \Delta \phi\right)-v^{}_\Delta v_\phi^2\right]+\frac{\lambda^{}_1}{4}\left\{[\tr\left(\Delta^{\dagger}_{}\Delta\right)]^{2}_{}-v_\Delta^4\right\}+\frac{\lambda^{}_2}{4}\left\{\tr\left[\left(\Delta^{\dagger}_{}\Delta\right)^{2}_{}\right]-v_\Delta^4\right\}\nonumber\\
	&+\lambda^{}_3\left[\left(\phi^\dagger_{} \phi\right)\tr\left(\Delta^\dagger_{} \Delta\right)-\frac{1}{2}v_\phi^2 v_\Delta^2\right]+\lambda^{}_4 \phi^\dagger_{} \Delta \Delta^\dagger_{} \phi\;.
\end{align}

With the above scalar potential, the total energy turns out to be
\begin{eqnarray}
	E(\mu)=\frac{4\pi v}{g}\int_0^{\infty} {\rm d}\xi \left({\cal H}^{}_{\rm gauge}+{\cal H}^{}_{\rm doublet}+{\cal H}^{}_{\rm triplet}\right)\;,
\end{eqnarray}
where ${\cal H}^{}_{\rm gauge}$ and ${\cal H}^{}_{\rm doublet}$ are the same as those in the minimal HTM [i.e., Eqs.~(\ref{eq:E1}) and (\ref{eq:E2})], and ${\cal H}^{}_{\rm triplet}$ is given by
\begin{align}
	\label{eq:Htrip}
	{\cal H}^{}_{\rm triplet}=&+\frac{\lambda_\Delta^2}{2g^2_{}\beta^2_{}}\xi^2_{}\left(2h^2_{}-1-h^{}_\Delta\right)\left(1-h^{}_\Delta\right)+\frac{v_\Delta^2}{6\beta v^2_\phi}\left[3\xi^2_{} h_\Delta'^2+16h_\Delta^2\left(1-f\right)^2_{}\sin^2_{}\mu\right]\nonumber\\
	&+\frac{\lambda_\Delta^2}{2g^2_{}\beta^2_{}}\xi^2_{}\left\{\frac{\kappa^2_{}-\left(\lambda-2\lambda_\Delta^2 \right)v_\phi^2}{\lambda_\Delta^2 v_\phi^2}\left(1-h^2_{}\right)\right.\nonumber\\
	&\left. +\left(\frac{v^{}_\Delta M^{}_\Delta}{\lambda^{}_\Delta v_\phi^2}-1\right)\left[2\left(1-h^2_{} h_\Delta^2\right)-\left(\frac{v^{}_\Delta M^{}_\Delta}{\lambda^{}_\Delta v_\phi^2}+1\right)\left(1-h_\Delta^2\right)\right]\right\}\nonumber\\
	&-\frac{\lambda^{}_1+\lambda^{}_2}{4g^2_{}\beta^2_{}}\frac{v_\Delta^4}{v_\phi^4}\xi^2_{}\left(1-h_\Delta^4\right)-\frac{\lambda^{}_3}{2g^2_{}\beta^2_{}}\frac{v_\Delta^2}{v_\phi^2}\xi^2_{}\left(1-h^2_{} h_\Delta^2\right)\;,
\end{align}
where $\beta$ is still defined as $\beta\equiv v^2_{}/v_\phi^2$.
Note that $\lambda_4$ does not appear in the energy, because $\phi^\dagger_{} \Delta \Delta^\dagger_{} \phi$ always vanishes with the ansatz in Eqs.~(\ref{eq:phi}) and (\ref{eq:Delta}). It is easy to check that in the limit of $\lambda^{}_1+\lambda^{}_2 =0$ and $\lambda^{}_3=0$, the parameters $\kappa^2_{}$ and $M^{}_\Delta$ are related to the VEVs by Eq.~(\ref{eq:vev}), then the 2nd to 4th lines of Eq.~(\ref{eq:Htrip}) vanish and Eq.~(\ref{eq:Htrip}) reduces to Eq.~(\ref{eq:E3}). Moreover, the terms in the 2nd to 4th lines of Eq.~(\ref{eq:Htrip}) are independent of the loop parameter $\mu$, implying that they do not influence the extreme points of the energy. Therefore, we conclude that the sphaleron configuration in the HTM with the full potential is still located at $\mu=\pi/2$.

In order to recast the sphaleron energy into a more compact form, we introduce the following dimensionless parameters
\begin{eqnarray}
	\label{eq:rhoparameter}
	\varrho^{}_{1}\equiv \frac{\lambda}{g^2_{}}\;,\quad
	\varrho^{}_{2}\equiv \frac{2\lambda_{\Delta}^2}{g^2_{}}\;,\quad
	\varrho^{}_{3}\equiv \frac{v_\Delta^2}{v_\phi^2}\;,\quad
	\varrho^{}_{4}\equiv \frac{\kappa^{2}_{}}{g^2_{} v_\phi^2}\;,\quad
	\varrho^{}_{5}\equiv \frac{M_{\Delta}^2}{g^2_{} v_\phi^2}\;.
\end{eqnarray}
Then $\lambda_{1}+\lambda_{2}$ and $\lambda_{3}$ are related to them via
\begin{eqnarray}
	\label{eq:relation}
	\lambda^{}_{1}+\lambda^{}_{2}=g^2_{}\left(-\frac{\varrho^{}_5}{\varrho^{}_{3}}-\frac{\varrho^{}_{5}}{\varrho^{}_{3}}\sqrt{\frac{\varrho^{}_{2}}{2\varrho^{}_{3}\varrho^{}_{5}}}+\frac{\varrho^{}_{1}-\varrho^{}_{4}}{\varrho_{3}^2}\right)\;,\quad
	\lambda^{}_3=g^2_{}\left(\frac{\varrho^{}_{4}-\varrho^{}_{1}}{\varrho^{}_{3}}+\sqrt{\frac{2\varrho^{}_{2}\varrho^{}_{5}}{\varrho^{}_{3}}}\right)\;.
\end{eqnarray} 
Notice that in the limit of $\lambda^{}_3 = 0$ and $\lambda^{}_1+\lambda^{}_2=0$, it goes back to the minimal HTM, where $\varrho^{}_4$ and $\varrho^{}_5$ are not independent and they are related to other three parameters by $\varrho^{}_4 =\varrho^{}_1-\varrho^{}_2$ and $\varrho^{}_5=\varrho^{}_2/(2\varrho^{}_3)$. With the help of Eq.~(\ref{eq:rhoparameter}), the sphaleron energy can be written as
\begin{align}
\label{eq:Esph-full}
	E^{}_{\rm sph}=\frac{4\pi v}{g}\int_{0}^{\infty}{\rm d}\xi & \left\{
	4f'^2_{}+\frac{8}{\xi^2_{}}f^2_{}\left(1-f\right)^2_{}+\frac{1}{\beta}\left(1-f\right)^2_{} h^2_{}+\frac{1}{2\beta}\xi^2_{} h'^2_{}\right.\nonumber\\
	&\left.+\frac{\xi^2}{4\beta^2_{}}\left[\left(\varrho^{}_1-\varrho^{}_2\right)\left(1-h^2_{}\right)^2_{}+\varrho^{}_2\left(h^2_{}-h^{}_\Delta\right)^2_{}\right]+\frac{\varrho^{}_3}{6\beta}\left[3\xi^2_{} h_\Delta'^2+16h_\Delta^2 \left(1-f\right)^2_{}\right]\right.\nonumber\\
	&\left.+\frac{\xi^2_{}}{4\beta^2_{}}\left[2\left(\varrho^{}_4-\varrho^{}_1+\varrho^{}_2 \right)\left(1-h^2_{}\right)-\left(2\varrho^{}_3 \varrho^{}_5 -\varrho^{}_2\right)\left(1-h_\Delta^2\right)\right]\right.\nonumber\\
	&\left.+\frac{\xi^2_{}}{2\beta^2_{}}\left(\sqrt{2\varrho^{}_2 \varrho^{}_3 \varrho^{}_5}-\varrho^{}_2\right)\left(1-h^2_{} h^{}_\Delta\right)+\frac{\xi^2_{}}{2\beta^2_{}}\left(\varrho^{}_1-\varrho^{}_4-\sqrt{2\varrho^{}_2 \varrho^{}_3 \varrho^{}_5}\right)\left(1-h^2_{} h_\Delta^2\right)\right.\nonumber\\
	&\left.-\frac{\xi^2_{}}{4\beta^2_{}}\left(\varrho^{}_1-\varrho^{}_4-\varrho^{}_3 \varrho^{}_5-\sqrt{\varrho^{}_2 \varrho^{}_3 \varrho^{}_5/2}\right)\left(1-h_\Delta^4\right)\right\}\;.
\end{align}
Starting with the energy, we obtain the sphaleron EOM via Eq.~(\ref{eq:EOMgeneral})
\begin{eqnarray}
	\xi^2_{} f'' &=& 2f\left(1-f\right)\left(1-2f\right)-\frac{\xi^2_{}}{4\beta}\left(1-f\right)h^2_{}-\frac{2\varrho^{}_3}{3\beta}\xi^2_{}\left(1-f\right)h_\Delta^2\;,\label{eq:EOM1-full}\\
	\left(\xi^2_{} h'\right)'&=&2\left(1-f\right)^2_{} h-\frac{\xi^2_{}}{\beta}\left[\left(\varrho^{}_1-\varrho^{}_2\right)h\left(1-h^2_{}\right)-\varrho^{}_2 h\left(h^2_{}-h^{}_\Delta\right)\right.\nonumber\\
	&&\left.+\left(\varrho^{}_4-\varrho^{}_1+\varrho^{}_2\right)h+\left(\sqrt{2\varrho^{}_2 \varrho^{}_3 \varrho^{}_5}-\varrho^{}_2\right)h h^{}_\Delta+\left(\varrho^{}_1-\varrho^{}_4-\sqrt{2\varrho^{}_2\varrho^{}_3 \varrho^{}_5}\right)h h_\Delta^2 \right]\;,\label{eq:EOM2-full}\\
	\varrho^{}_3 \left(\xi^2_{} h_\Delta'\right)' &=& \frac{16}{3}\varrho^{}_3 \left(1-f\right)^2_{} h^{}_\Delta -\frac{\varrho^{}_2 \xi^2_{}}{2\beta} \left(h^2_{}-h^{}_\Delta\right)+\frac{\xi^2_{}}{2\beta}\left[\left(2\varrho^{}_3 \varrho^{}_5 - \varrho^{}_2 \right)h^{}_\Delta-\left(\sqrt{2\varrho^{}_2 \varrho^{}_3 \varrho^{}_5}-\varrho^{}_2\right)h^2_{}\right.\nonumber\\
	&&\left.-2\left(\varrho^{}_1-\varrho^{}_4-\sqrt{2\varrho^{}_2 \varrho^{}_3 \varrho^{}_5}\right) h^2_{} h^{}_\Delta+2\left(\varrho^{}_1-\varrho^{}_4-\varrho^{}_3 \varrho^{}_5-\sqrt{\varrho^{}_2 \varrho^{}_3 \varrho^{}_5/2}\right)h_\Delta^3\right]\;.\label{eq:EOM3-full}
\end{eqnarray}
The profile functions $f$, $h$ and $h_\Delta$ should also satisfy the boundary conditions in Eq.~(\ref{eq:BC}).

Although there are totally 8 parameters in the scalar potential, namely $\lambda$, $\lambda^{}_{\Delta}$, $\kappa^2_{}$, $M^{}_\Delta$, and $\lambda^{}_i$ (for $i=1,2,3,4$), the sphaleron configuration is only affected by 5 independent parameters, i.e., $\varrho^{}_{1}$-$\varrho^{}_{5}$ defined in Eq.~(\ref{eq:rhoparameter}). This implies that not all parameters in the HTM are relevant to the $B$-violating process.

\subsection{Constraints on the Parameters}
\label{subsec:constraints}
We have seen that the sphaleron configuration in the HTM is determined by 5 parameters. Using the spectral method developed in Appendix~\ref{app:spectral}, one can solve Eqs.~(\ref{eq:EOM1-full})-(\ref{eq:EOM3-full}) and calculate the sphaleron energy in Eq.~(\ref{eq:Esph-full}) for any given parameters. However, there are constraints from both theoretical and experimental aspects on the parameters in the HTM~\cite{Dev:2013xol,Dev:2017ouk,Dev:2018sel,Dev:2018tox, Primulando:2019evb,Ashanujjaman:2021txz,Mandal:2022zmy}. Below we list all the constraints that are relevant to the sphaleron.
\begin{itemize}
\item Triplet VEV: From the first equality of Eq.~(\ref{eq:relation}) one can obtain
\begin{eqnarray}
	\label{eq:rho4appro}
	\varrho^{}_{4}&=&\varrho^{}_1-\frac{1}{2}\varrho^{}_{3}\varrho^{}_{5}\left(2+\sqrt{\frac{2\varrho^{}_{2}}{\varrho^{}_{3}\varrho^{}_{5}}}\right)-\frac{\left(\lambda^{}_{1}+\lambda^{}_{2}\right)\varrho_{3}^2}{g^2_{}}\nonumber\\
	&\approx& \varrho^{}_1-\frac{1}{2}\varrho^{}_{3}\varrho^{}_{5}\left(2+\sqrt{\frac{2\varrho^{}_{2}}{\varrho^{}_{3}\varrho^{}_{5}}}\right)\;,
\end{eqnarray}
where in the second line we have neglected the term proportional to $\varrho_{3}^2$. This is a good approximation because the EW precision measurements require $\varrho^{}_{3}\lesssim 10^{-3}$, and $\lambda^{}_{i}$ cannot be too large for unitarity. Therefore, $\varrho^{}_{4}$ can be approximated using Eq.~(\ref{eq:rho4appro}) in the calculation of the sphaleron. Substituting Eq.~(\ref{eq:rho4appro}) back to the second equality of Eq.~(\ref{eq:relation}) we have
\begin{eqnarray}
	\label{eq:lambda3appro}
	\frac{\lambda^{}_{3}}{g^2}\approx \sqrt{\frac{\varrho^{}_{2}\varrho^{}_{5}}{2\varrho^{}_{3}}}-\varrho^{}_{5}\;.
\end{eqnarray}

\item Bounded-from-below conditions and the requirement of unitarity: These conditions provide a series of inequalities on the couplings $\lambda_i$ in the scalar potential, and part of them can be translated to the constrains on $\rho_i$. For a complete set of these constraints, see Refs.~\cite{Arhrib:2011uy,Bonilla:2015eha}. Here we only list those which are relevant to the sphaleron:
\begin{align}
	\label{eq:BFB-unitarity}
0 < \varrho^{}_{1} \leqslant \frac{4\pi}{g^2_{}}\;,\quad
-\sqrt{\frac{4\pi}{g^2_{}} \varrho^{}_{1}} < \sqrt{\frac{\varrho^{}_{2}\varrho^{}_{5}}{2\varrho^{}_{3}}}-\varrho^{}_{5}\leqslant \frac{4\pi}{g^2_{}}\;,\quad
\varrho^{}_{1}-\varrho^{}_{3}\varrho^{}_{5}-\sqrt{\varrho^{}_{2}\varrho^{}_{3}\varrho^{}_{5}/2}>0\;.
\end{align}
In addition, there are also constraints relevant to $\lambda_{4}$:
\begin{eqnarray}
	\label{eq:lambda34}
-\sqrt{\frac{4\pi}{g^2_{}}\varrho^{}_{1}}<\frac{\lambda^{}_3+\lambda^{}_{4}}{g^2_{}}\leqslant \frac{4\pi}{g^2_{}}\;,\quad
\left|2\lambda^{}_3 + 3\lambda^{}_{4}\right|\leqslant8\pi\;,\quad
\left|2\lambda^{}_{3}-\lambda^{}_{4}\right|\leqslant 8\pi\;.
\end{eqnarray}
Although $\lambda^{}_{4}$ does not directly contribute to the sphaleron configuration, it would be related to other parameters via the Higgs mass (as discussed below). 

\item Higgs mass: The HTM should also predict a CP-even neutral Higgs boson $h$, whose mass is around 125~GeV. In the HTM, the mass of $h$ is predicted by
\begin{align}
	\label{eq:Higgsmass}
m_h^2=g^2_{} v_\phi^2 &\left[\varrho^{}_{1}+\frac{1}{2}\sqrt{\frac{\varrho^{}_{2}\varrho^{}_{5}}{2\varrho^{}_{3}}}+\frac{\lambda^{}_{1}+\lambda^{}_{2}}{g^2_{}}\varrho^{}_{3}\right.\nonumber\\
&\left.-\sqrt{\left(\varrho^{}_{1}-\frac{1}{2}\sqrt{\frac{\varrho^{}_{2}\varrho^{}_{5}}{2\varrho^{}_{3}}}-\frac{\lambda^{}_{1}+\lambda^{}_{2}}{g^2_{}}\varrho^{}_{3}\right)^2_{}+4\left(\sqrt{\varrho^{}_{2}\varrho^{}_{5}/2}-\frac{\lambda^{}_{3}+\lambda^{}_{4}}{g^2_{}}\sqrt{\varrho^{}_{3}}\right)^2_{}}\right]\;.
\end{align}
The terms proportional to $\lambda^{}_1+\lambda^{}_2$ in Eq.~(\ref{eq:Higgsmass}) are suppressed by $\varrho^{}_{3}$ and can be safely neglected. Then one can extract $\lambda^{}_4$ in terms of $m_h$ and $\varrho^{}_i$:
\begin{eqnarray}
	\label{eq:lambda4}
	\frac{\lambda^{}_4}{g^2_{}}\approx \varrho^{}_{5} \pm \frac{1}{\left(2\varrho^{}_{3}\right)^{3/4}}\sqrt{\left(\varrho^{}_{1}-\frac{m_h^2}{2 g^2_{} v_\phi^2}\right)\left(\sqrt{\varrho^{}_{2}\varrho^{}_{5}}-\frac{\sqrt{2\varrho^{}_{3}}m_h^2}{g^2_{} v_\phi^2}\right)}\;.
\end{eqnarray}
Given $g\approx 0.65$, $m^{}_h\approx 125~{\rm GeV}$ and $v^{}_\phi\approx 246~{\rm GeV}$, the combination of Eqs.~(\ref{eq:lambda4}), (\ref{eq:lambda3appro}) and (\ref{eq:lambda34}) provides additional constraints on $\varrho^{}_i$.

\item Collider constraints:
The collider searches put the lower bound on the mass of doubly-charged Higgs, namely  $m^{}_{H^{\pm\pm}_{}}\gtrsim 350~{\rm GeV}$ or $m^{}_{H^{\pm\pm}_{}}\gtrsim 1~{\rm TeV}$ for the decay channels dominated by vector-boson ($v^{}_\Delta \gtrsim 10^{-4}_{}~{\rm GeV}$) or charged-lepton ($v^{}_\Delta \lesssim 10^{-4}_{}~{\rm GeV}$) final states, respectively~\cite{ATLAS:2021jol,ATLAS:2022pbd}. In the HTM, the mass of the doubly-charged Higgs is predicted to be
\begin{eqnarray}
	\label{eq:mH++}
m_{H^{\pm\pm}_{}}^2=g^2_{}v_\phi^2\left(\sqrt{\frac{\varrho^{}_{2}\varrho^{}_{5}}{2\varrho^{}_{3}}}-\frac{\lambda^{}_{4}}{g^2_{}}-\frac{\lambda^{}_{2}}{g^2_{}}\varrho^{}_{3}\right)\approx g^2_{}v_\phi^2\left(\sqrt{\frac{\varrho^{}_{2}\varrho^{}_{5}}{2\varrho^{}_{3}}}-\frac{\lambda^{}_{4}}{g^2_{}}\right)\;.
\end{eqnarray}
For $\varrho^{}_{3}=10^{-3}_{}$, the dominant decay channel is the gauge-boson final state, so the collider constraint implies
\begin{eqnarray}
	\label{eq:doubly-charged}
	\sqrt{\frac{\varrho^{}_{2}\varrho^{}_{5}}{2\varrho^{}_{3}}}-\frac{\lambda^{}_{4}}{g^2_{}}\gtrsim 4.8\;,
\end{eqnarray}
where $g\approx0.65$ and $v^{}_\phi\approx 246~{\rm GeV}$ have been used.

\item Charged lepton flavor violation (cLFV): The lack of the observation of cLFV in the HTM gives~\cite{Dinh:2012bp}
\begin{eqnarray}
	\label{eq:rho3rho5}
M^{}_\Delta v^{}_\Delta  \gtrsim 10^{2}~{\rm GeV}\cdot{\rm eV} \quad \Rightarrow \quad \varrho^{}_{3}\varrho^{}_{5} \gtrsim 10^{-24}_{}\;.
\end{eqnarray}
This constraint is easy to satisfy for $v^{}_\Delta \sim {\cal O}(\rm GeV)$.
\end{itemize}

In summary, the relevant constraints on the parameters that contribute to the sphaleron configuration are given by Eqs.~(\ref{eq:BFB-unitarity}), (\ref{eq:lambda34}), (\ref{eq:doubly-charged}) and (\ref{eq:rho3rho5}), where $\lambda^{}_{3}$ and $\lambda^{}_{4}$ are given by Eqs.~(\ref{eq:lambda3appro}) and (\ref{eq:lambda4}), respectively.

\subsection{Sphaleron Solution}
\begin{figure*}[t!]
	\centering
	\includegraphics[scale=0.8]{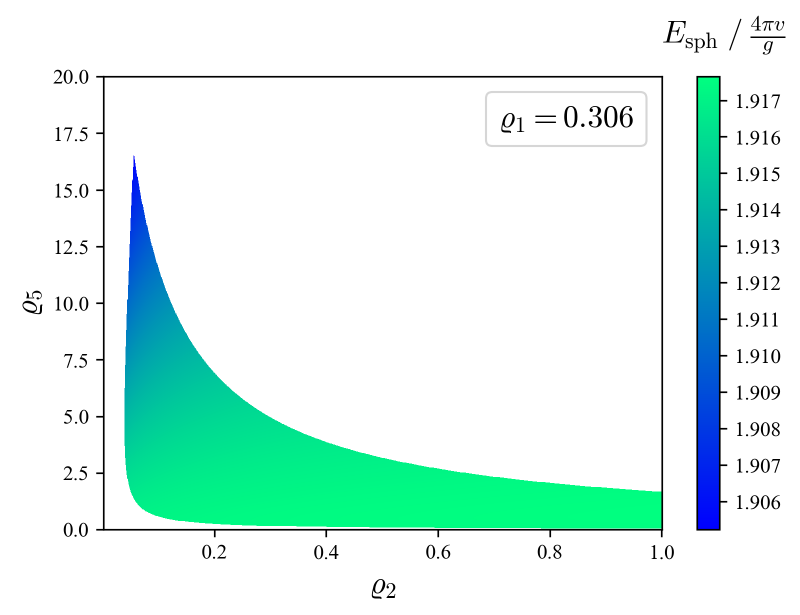}
	\caption{\label{fig:energy25}The sphaleron energy in the HTM with respect to the doublet-triplet trilinear coupling $\varrho^{}_{2}$ and the triplet mass parameter $\varrho^{}_{5}$, where $\varrho^{}_{1}=0.306$ and $\varrho^{}_{3}=10^{-3}_{}$ are fixed. Moreover, $\varrho^{}_{4}$ is calculated by Eq.~(\ref{eq:rho4appro}).}
\end{figure*}

\begin{figure*}[t!]
	\centering
	\includegraphics[scale=0.63]{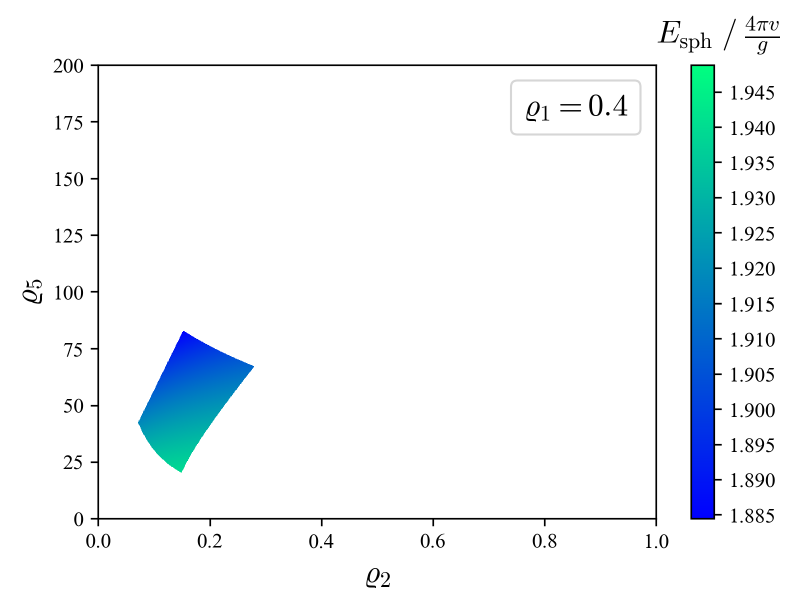}\quad
	\includegraphics[scale=0.63]{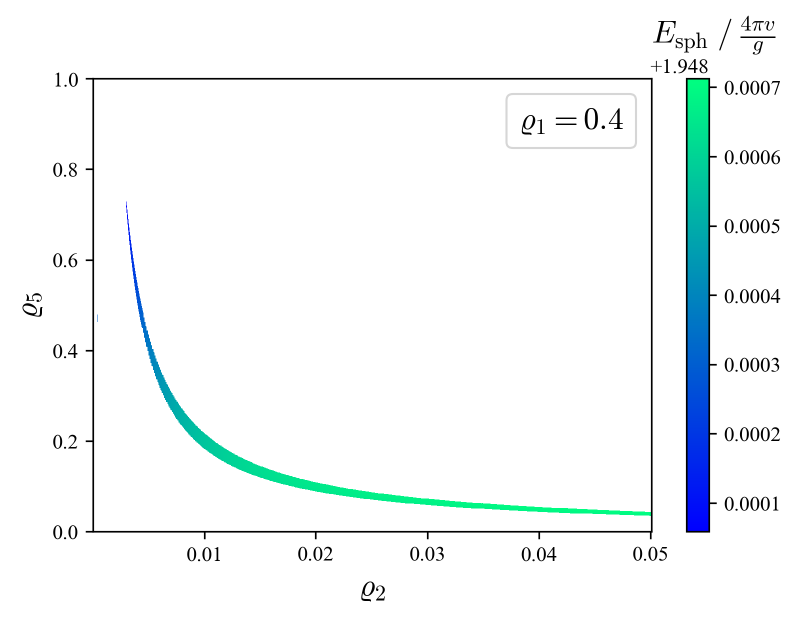}\quad\\
	\vspace{0.2cm}
	\includegraphics[scale=0.63]{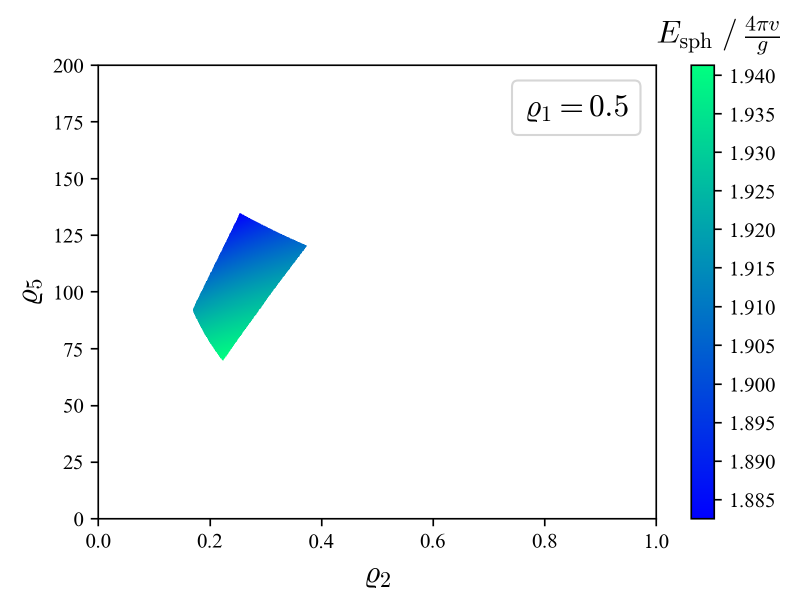}\quad
	\includegraphics[scale=0.63]{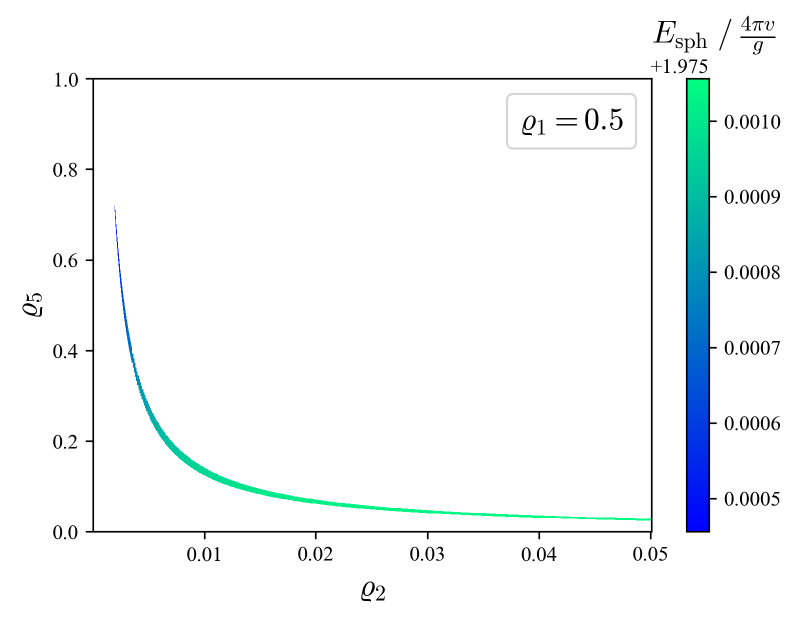}\\
	\vspace{0.2cm}
	\includegraphics[scale=0.63]{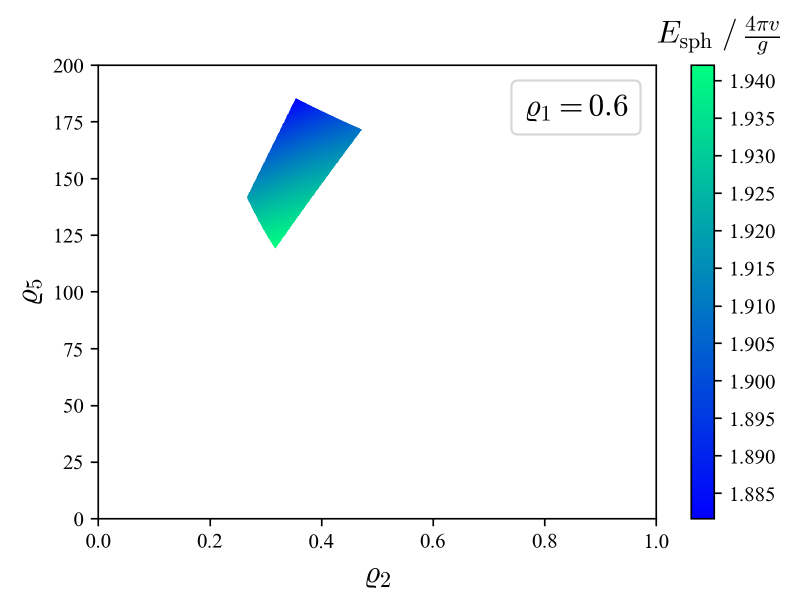}\quad
	\includegraphics[scale=0.63]{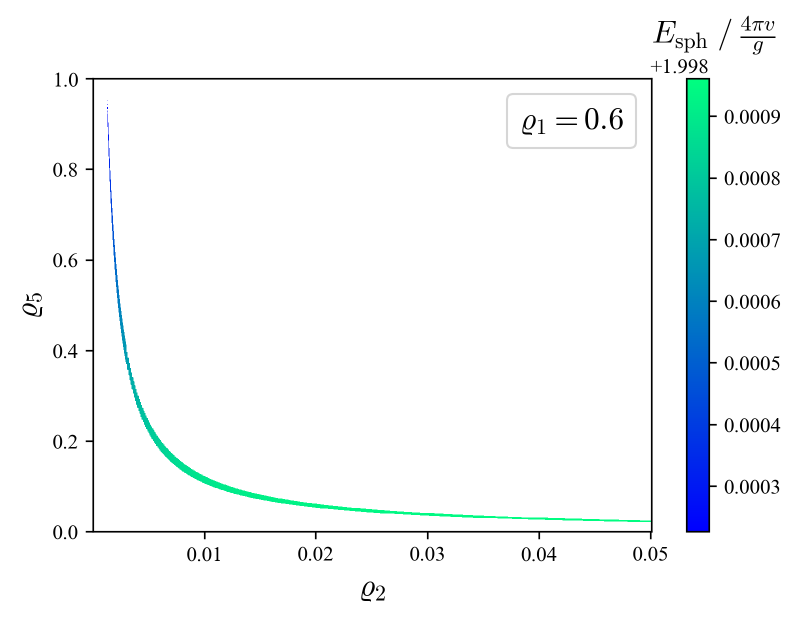}\\
	\vspace{0.2cm}
	\caption{\label{fig:energ25extra} The sphaleron energy in the HTM with respect to $\varrho^{}_{2}$ and $\varrho^{}_{5}$, where $\varrho^{}_{3}=10^{-3}$ is fixed, and $\varrho^{}_{1}$ is taken to be $0.4$, $0.5$, and $0.6$, respectively. Moreover, $\varrho^{}_{4}$ is calculated via Eq.~(\ref{eq:rho4appro}). When $\varrho^{}_{1}\gtrsim 0.34$, the allowed parameter space of $\varrho^{}_{2}$ and $\varrho^{}_{5}$ begins to split into two distinct regions. This is due to the constraints considered in Sec.~\ref{subsec:constraints}. {\bf Region A} (left panel) corresponds to relatively large values of $\varrho^{}_{2} \varrho^{}_{5}$, while the allowed values of $\varrho^{}_{2}\varrho^{}_{5}$ in {\bf Region B}  (right panel) is much smaller.} 
\end{figure*}

\begin{figure*}[t!]
	\centering
	\includegraphics[scale=0.63]{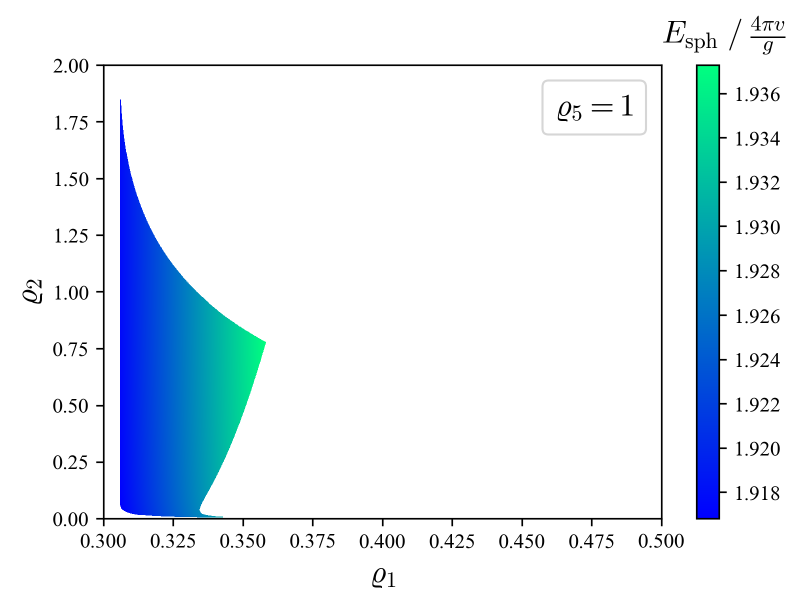}\quad
	\includegraphics[scale=0.63]{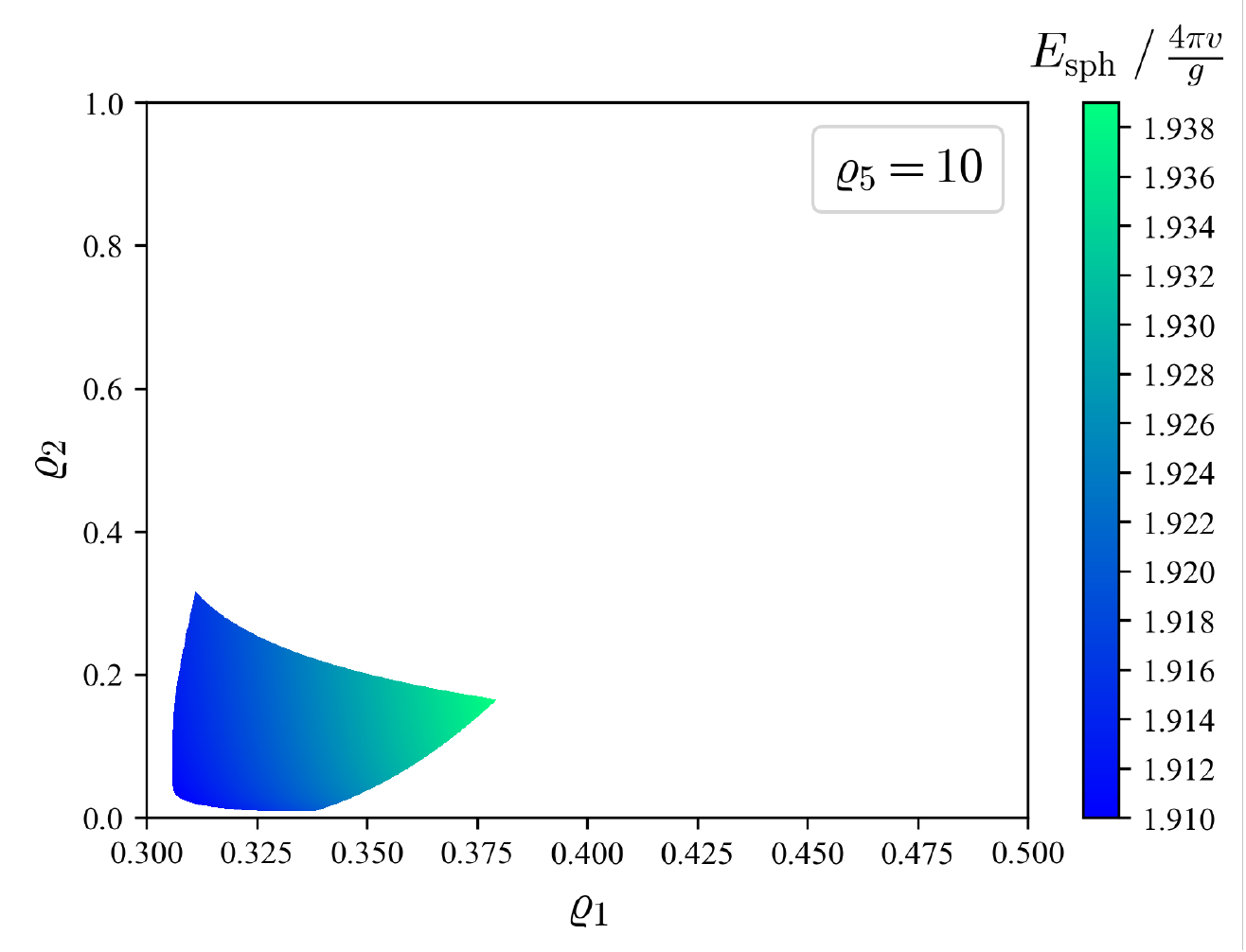}\quad\\
	\vspace{0.2cm}
	\includegraphics[scale=0.63]{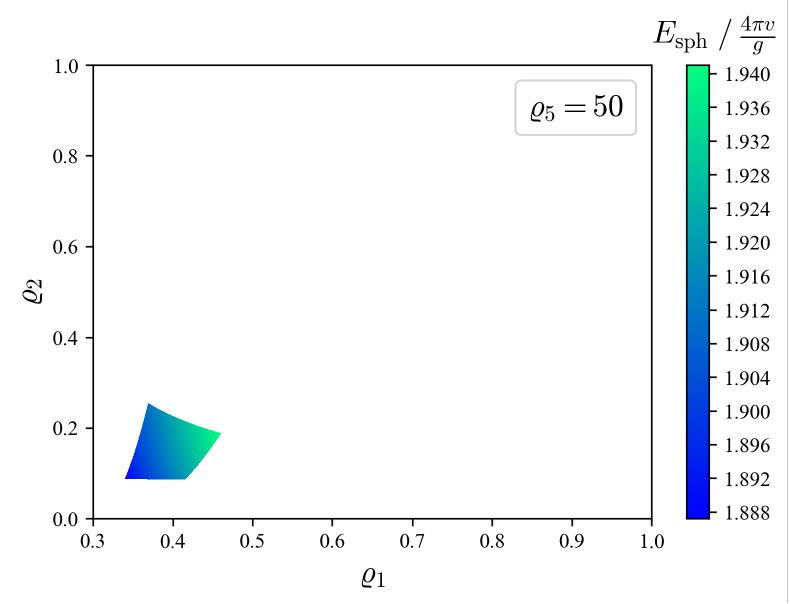}\quad
	\includegraphics[scale=0.63]{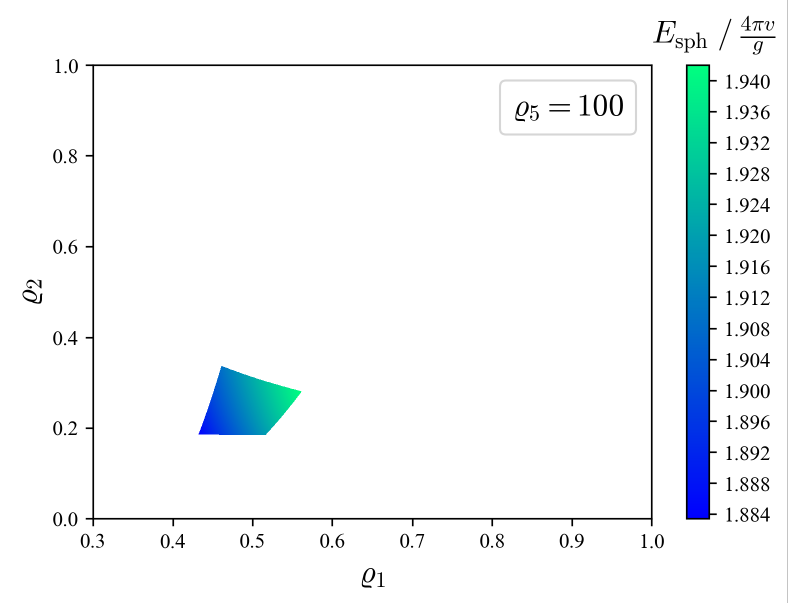}\\
	\vspace{0.2cm}
	\includegraphics[scale=0.63]{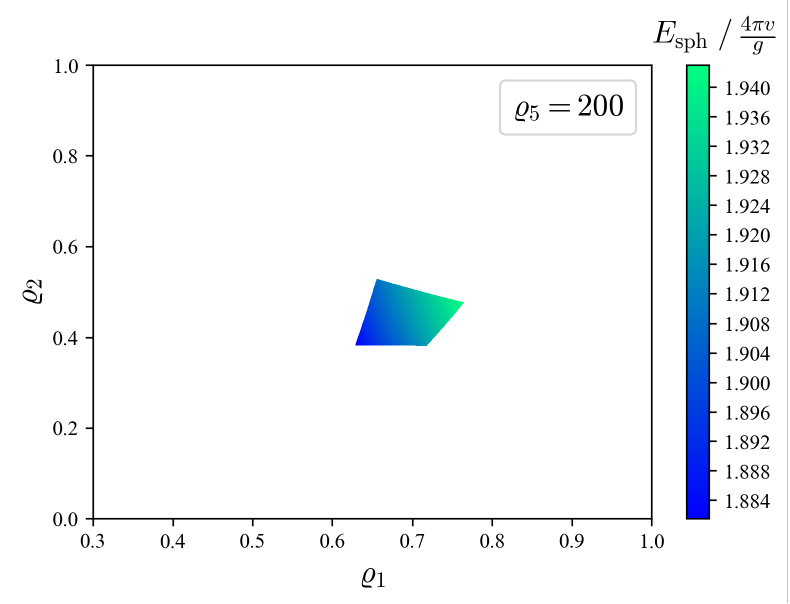}\quad
	\includegraphics[scale=0.63]{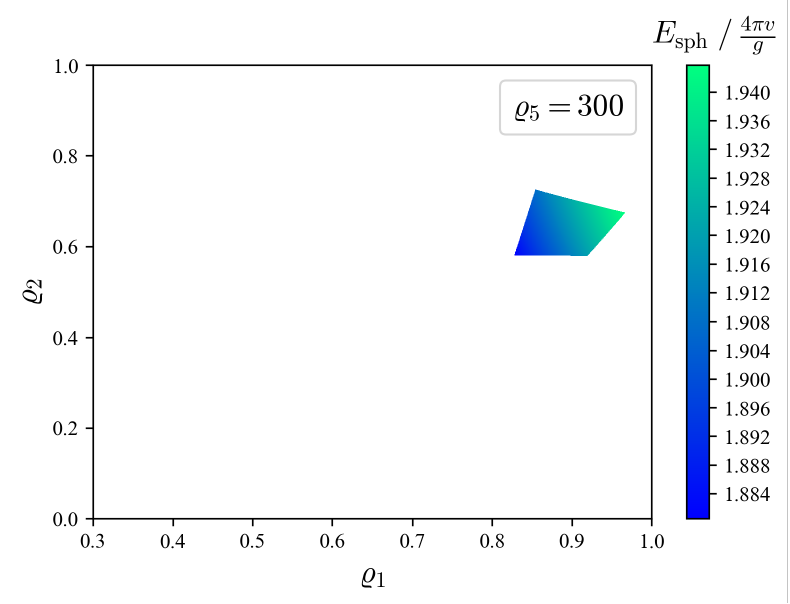}\\
	\vspace{0.2cm}
	\caption{\label{fig:energy12}The sphaleron energy in the HTM with respect to the doublet quartic coupling $\varrho^{}_{1}$ and the doublet-triplet trilinear coupling $\varrho^{}_{2}$, where $\varrho^{}_{3}=10^{-3}_{}$ is fixed, and $\varrho^{}_{5}$ is taken to be 1, 5, 50, 100, 200, 300, respectively. Moreover, $\varrho^{}_{4}$ is calculated from Eq.~(\ref{eq:rho4appro}).} 
\end{figure*}
Basically, the contribution of the triplet to the sphaleron energy is suppressed by its VEV. For a small enough VEV-ratio parameter $\varrho^{}_{3}$, it should reduce to the SM case. Therefore, we fix $\varrho^{}_{3}$ to be its upper bound (i.e., $\varrho^{}_{3}=10^{-3}_{}$), and see how much the difference of the sphaleron energy between the HTM and the SM is under all theoretical and experimental constraints. In addition, $\varrho^{}_{4}$ could be calculated from Eq.~(\ref{eq:rho4appro}) as a good approximation. Therefore, we are left with three independent parameters, namely the doublet quartic coupling $\varrho^{}_{1}$, the doublet-triplet trilinear coupling $\varrho^{}_{2}$, and the triplet mass parameter $\varrho^{}_{5}$.

In the SM, $\varrho^{}_{1}$ is completely fixed by the Higgs mass, i.e., $\varrho_{1}^{\rm SM}=m_h^2/\left(2g^2_{} v_\phi^2\right)\approx 0.306$, and so is the sphaleron energy $E_{\rm sph}^{\rm SM}\approx 1.92\times 4\pi v/g$. However, in the HTM, $\varrho^{}_{1}$ is not fixed because the Higgs mass depends on other parameters [see Eq.~(\ref{eq:Higgsmass})]. It is not difficult to prove that  for $\varrho^{}_{1}<\varrho_{1}^{\rm SM}$ there is no allowed parameter space under the constraints discussed in Sec.~\ref{subsec:constraints}. Therefore we must have $\varrho^{}_{1}\geqslant \varrho_{1}^{\rm SM}\approx 0.306$ and $\varrho^{}_{2} \varrho^{}_{5} \geqslant  2 \varrho^{}_{3} m_h^4/\left(g^4_{} v_\phi^4\right)\approx 7.5 \times 10^{-4}_{}$. 
In Fig.~\ref{fig:energy25}, we have taken $\varrho^{}_{1}=0.306$ and shown the sphaleron energy with respect to $\varrho^{}_{2}$ and $\varrho_{5}$. It is clear that a larger $\varrho^{}_{5}$ (corresponding to a heavier triplet) would decrease the sphaleron energy, though the difference is small compared with the SM case because of the suppression from $\varrho^{}_{3}$. 

However, unlike the SM where $\varrho^{}_1$ is fixed to be 0.306, $\varrho^{}_{1}>0.306$ is also allowed in the HTM. Due to the constraints in Sec.~\ref{subsec:constraints}, the parameter space of $\varrho_{2}$ and $\varrho^{}_{5}$ begins to split into two distinct regions when $\varrho^{}_{1}\gtrsim 0.34$, as is shown in Fig.~\ref{fig:energ25extra}. 
In {\bf Region A} (left panel of Fig.~\ref{fig:energ25extra}), it can be seen that the allowed parameter space of $\varrho^{}_{2}$ and $\varrho^{}_{5}$ moves to upper-right as $\varrho^{}_{1}$ increases. This can be understood by observing the expression of $\lambda^{}_{4}$ in Eq.~(\ref{eq:lambda4}), whose magnitude should be bounded by the requirement of unitarity. Moreover, the sphaleron energy decreases as $\varrho^{}_{5}$ increases, while larger $\varrho^{}_{1}$ would bring about larger sphaleron energies. 
The value of $\varrho^{}_{1}$ can keep increasing until the unitarity bound, i.e., $\varrho_{1}^{\rm max}=4\pi/g^2$, is reached. We have verified numerically that the maximum sphaleron energy in {\bf Region A} is around $1.97\times 4\pi v/g$. Basically, the parameters in {\bf Region A} correspond to a heavy mass scale $M^{}_\Delta$ of the triplet scalar, which can reach TeV or above.

Things are quite different for {\bf Region B} (shown in the right panel of Fig.~\ref{fig:energ25extra}). The allowed values of $\varrho^{}_{2}$ and $\varrho^{}_{5}$ are much smaller. More explicitly, the lower and upper bounds of $\varrho^{}_{2}\varrho^{}_{5}$ in {\bf Region B} are given by
\begin{eqnarray}
\label{eq:narrowbound}
     \sqrt{\varrho^{}_2\varrho^{}_5} &\leqslant& \frac{1}{2}\left[\frac{1}{\sqrt{2\varrho^{}_3}}\left(\varrho^{}_1-\frac{m_h^2}{2 g^2_{} v^2_\phi}\right)-\frac{4\sqrt{2\pi}}{g}\sqrt{\varrho^{}_1\varrho^{}_3}-\sqrt{A^{}_1}\right]\;,\nonumber\\
    \sqrt{\varrho^{}_2\varrho^{}_5} &\geqslant& \frac{1}{2}\left[2\sqrt{2\varrho^{}_3}\left(\varrho^{}_5+\frac{24}{5}\right)+\frac{1}{\sqrt{2\varrho^{}_3}}\left(\varrho^{}_1-\frac{m_h^2}{2g^2_{}v^2_\phi}\right)-\sqrt{A^{}_2}\right]\;,
    \end{eqnarray}
where 
\begin{eqnarray}
    A^{}_1 &\equiv& \frac{1}{2\varrho^{}_3}\left(\varrho^{}_1-\frac{m_h^2}{2 g^2_{} v^2_\phi}\right)\left[\varrho^{}_1-\frac{16\sqrt{\pi}\varrho^{}_3}{g}\sqrt{\varrho^{}_1}-\frac{m^2_h}{2g^2_{} v^2_\phi}\left(1+16\varrho^{}_3\right)\right]\;,\nonumber\\
    A^{}_2 &\equiv& \frac{1}{10\varrho^{}_3}\left(\varrho^{}_1-\frac{m^2_h}{2g^2_{}v^2_\phi}\right)\left[5\varrho^{}_1+8\varrho^{}_3\left(24+5\varrho^{}_5\right)-\frac{5 m_h^2}{2g^2{} v^2_\phi}\left(1+16\varrho^{}_3\right)\right]\;.\nonumber
\end{eqnarray}
Note that Eq.~(\ref{eq:narrowbound}) comes from the constraints in Sec.~\ref{subsec:constraints}, where $m^{}_h\approx 125~{\rm GeV}$, $g\approx 0.65$ and $\varrho^{}_{3}=10^{-3}_{}$ should be substituted to evaluate the lower and upper bounds. The allowed  values of $\varrho^{}_{2}$ and $\varrho^{}_{5}$ are restricted to a narrow parameter space by Eq.~(\ref{eq:narrowbound}). 
For example, for $\varrho^{}_{1}=0.6$, the validity of Eq.~(\ref{eq:narrowbound}) requires $\varrho^{}_{5}\lesssim 0.987$ and $1.05 \times 10^{-3}_{} \lesssim \varrho^{}_{2}\varrho^{}_{5} \lesssim 1.22\times 10^{-3}_{}$, which corresponds to the narrow band in the bottom-right subfigure of Fig.~\ref{fig:energ25extra}. Since $\varrho^{}_{5}$ is relatively small, the sphaleron energy in {\bf Region B} can be significantly enhanced as $\varrho^{}_{1}$ increases. In particular, for $\varrho^{}_{1}=\varrho_{1}^{\rm max}=4\pi/g^2$, the sphaleron energy can reach $2.48\times 4\pi v/g$, which is enhanced by about $30\%$ compared with the sphaleron energy in the SM. The parameters in {\bf Region B} correspond to a much smaller $M^{}_\Delta$ than that in {\bf Region A} (basically lighter than 1~{\rm TeV}). However, it does not violate the collider constraints on the mass of doubly-charged Higgs, because $m^{}_{H^{\pm \pm}}$ depends on the combination of $\varrho^{}_{2}\varrho^{}_{5}$ rather than $\varrho^{}_5$ itself, and is enhanced by $\varrho_{3}^{-1/2}$ [see Eq.~(\ref{eq:mH++})]. On the other hand, since the allowed parameter space  in {\bf Region B} is quite narrow and is sensitive to the lower bound of $m^{}_{H^{\pm \pm}}$, we point out that it is readily testable by future collider searches and EW precision measurements.

In Fig.~\ref{fig:energy12}, we have shown the sphaleron energy with respect to $\varrho^{}_{1}$ and $\varrho^{}_{2}$ for different values of $\varrho^{}_{5}$. Note that all allowed parameters in Fig.~\ref{fig:energy12} belong to {\bf Region A} because the corresponding values of $\varrho^{}_{5}$ are not small enough to satisfy Eq.~(\ref{eq:narrowbound}). It is clear that for larger $\varrho^{}_{5}$, the allowed parameter space moves to upper-right. The increase of $\varrho^{}_{1}$ (or $\varrho^{}_{5}$) would enhance (or reduce) the sphaleron energy. For $\varrho^{}_{5}\gtrsim 100$ (corresponding to $M^{}_\Delta \gtrsim 1.6~{\rm TeV}$), the lower bound of the sphaleron energy tends to about $1.88\times 4\pi v/g$.

To sum up, the sphaleron energy in the SM is completely fixed by the Higgs mass, while that in the HTM is not. The allowed parameter space begins to split into two regions when $\varrho^{}_{1}\gtrsim 0.34$. In {\bf Region A}, the sphaleron energy is bounded to be $1.88 \times 4\pi v/g \lesssim E_{\rm sph}^{} \lesssim 1.97 \times 4 \pi v/g$. The difference of the sphaleron energy between the HTM and the SM is less than 3\%. On the contrary, in {\bf Region B}, since $\varrho^{}_{5}$ is relatively small, the sphaleron energy could be significantly enhanced as $\varrho^{}_{1}$ increases. Therefore we have $1.92 \times 4\pi v/g \lesssim E_{\rm sph}^{} \lesssim 2.48 \times 4 \pi v/g$, where the sphaleron energy could be enhanced up to about 30\% compared with the SM case.

\section{Summary and Discussions}
\label{sec:summary}
The origin of neutrino masses and the baryon asymmetry of the Universe are two of the most important unsolved problems in the SM. Both of them are possible to be explained in a unified framework of the HTM, which extends the SM by adding a complex triplet scalar. The couplings of the triplet to the gauge fields and to the SM Higgs field are expected to affect the sphaleron configuration in the SM, which plays an important role in baryogenesis. Therefore, to realize a self-consistent baryogenesis in the HTM, either via EW baryogenesis or via leptogenesis, the calculation of the sphaleron energy is indispensable.

In this work, we calculate the sphaleron configuration in the HTM for the first time, where both the doublet and the triplet scalar fields exist. Although there are 8 parameters in the scalar potential of the HTM, we find that the sphaleron configuration is determined by only 5 independent parameters, i.e., those defined in Eq.~(\ref{eq:rhoparameter}). Among them, the doublet quartic parameter $\varrho^{}_{1}$ would increase the sphaleron energy, as in the SM case; while the doublet-triplet trilinear parameter $\varrho^{}_{2}$, the VEV-ratio parameter $\varrho^{}_{3}$, and the triplet mass parameter $\varrho^{}_{5}$ would decrease the sphaleron energy in general compared with the SM. Nevertheless, at zero temperature, the constraint from EW precision measurements on the triplet VEV puts a stringent upper bound on $\varrho^{}_{3}$, thus highly suppresses the difference of the sphaleron energy between the HTM and the SM. Interestingly, we find there still exists some narrow parameter space where the sphaleron energy could be enhanced by 30\% compared with the SM case. Such narrow parameter space can be tested by future collider searches of doubly-charged Higgs and EW precision measurements. The enhancement of the sphaleron energy by 30\%  may have a great impact on the sphaleron rate in the broken phase, which is proportional to ${\rm exp}\left(-E^{}_{\rm sph}/T\right)$. More explicitly, a larger sphaleron energy leads to a smaller sphaleron rate. When the sphaleron rate in the broken phase becomes smaller than the Hubble expansion rate, the departure from thermal equilibrium occurs and thus it is more difficult to washout the baryon number asymmetry in the HTM than in the SM. Nevertheless, in order to perform a full study of electroweak baryogenesis in the HTM, a more dedicated calculation of the sphaleron rate is desirable.

In the following, we discuss some possible extensions of the present work.
All of the calculations in this paper have neglected the finite-temperature effects. However, the sphaleron transition rate is significant above the temperature of ${\cal O}(100)~{\rm GeV}$ in the early Universe, which is a crucial process for baryogenesis. Therefore, in principle one should include the finite-temperature corrections as well as the one-loop corrections into the scalar potential in Eq.~(\ref{eq:fullpotential}) and recalculate the sphaleron configuration using the formalism developed above. This is beyond the scope of this paper, and will be left for a future work. As a good approximation, one could estimate the sphaleron energy at finite temperatures using the scaling law~\cite{Braibant:1993is,Moreno:1996zm}
\begin{eqnarray}
E^{}_{\rm sph}(T)=E^{}_{\rm sph}\frac{v(T)}{v}\;,
\end{eqnarray}
where $v$ and $E_{\rm sph}$ are the VEV and the sphaleron energy at zero temperature, and $v(T)=\left[v_\phi^2(T)+2v_\Delta^2(T)\right]^{1/2}_{}$ is the VEV at a finite temperature, with $v^{}_\phi(T)$ and $v^{}_\Delta(T)$ being the VEVs of the doublet and the triplet. On this point, it is worthwhile to emphasize that $v^{}_\Delta(T)/v^{}_\phi(T)$ is not constrained by experiments as at zero temperature, and hopefully we could have a larger $\varrho^{}_{3}$ at finite temperatures. As has been shown in the right panel of Fig.~\ref{fig:energy-HTM-min}, a large $\varrho^{}_{3}$ would significantly decrease the sphaleron energy compared with the SM. 

Apart from the finite-temperature effects, one can study the sphaleron configuration in the Georgi-Machacek (GM) model~\cite{Georgi:1985nv,Chanowitz:1985ug,Gunion:1989ci}. The GM model further extends the HTM by introducing an additional real triplet scalar with hypercharge $Y=0$, and can maintain the custodial symmetry at the tree level by adjusting the VEVs of the complex and real triplets. In this way, the VEVs of the triplets are no longer suppressed and can even be larger than that of the doublet. This may significantly change the sphaleron configuration in the SM according to the results in this work. Therefore, it would be interesting to calculate the sphaleron energy and investigate whether a successful EW baryogenesis could be carried out in the GM model, given that the strong first-order EW phase transition is possible in this model~\cite{Chiang:2014hia,Zhou:2018zli}.


Another interesting extension of this work is to study the two-step EW phase transition in the HTM (see Ref.~\cite{Wu:2023mjb} for a recent discussion on this point). In such a scenario, the triplet VEV $v^{}_\Delta$ at the end of the first-step phase transition is not constrained by EW precision measurements, so it is possible to have $\varrho^{}_{3} \gg 1$ at that time. According to the results in this work, the sphaleron energy can be significantly reduced when compared with the SM case.

Finally, it is worthwhile to mention that although the classical sphaleron configuration is obtained as the saddle-point solution of the energy functional, quantum corrections to the sphaleron energy may also be important. The calculation of quantum corrections to the sphaleron energy, as far as we know, is still lacking even in the SM case. This will be left for future works.

\vspace{0.5cm}

\noindent{\bf Note added.} During the final preparation of this paper, a relevant work~\cite{Wu:2023mjb} appeared, which studied the sphaleron configuration in extensions of the SM with general electroweak multiplets (see also Ref.~\cite{Ahriche:2014jna} for earlier efforts). In particular, Ref.~\cite{Wu:2023mjb} calculated the sphaleron energy in a septuplet extension of the SM. Besides, Ref.~\cite{Wu:2023mjb} focused on the scenario where the neutral component of the multiplet can be a dark matter candidate. In this case, the hypercharge of the multiplet should be zero and the VEV is vanishing at zero temperature. This is different from the scenario we considered in the current work.

\section*{Acknowledgements}
We would like to thank Huai-Ke Guo, Yu Tian, Yanda Wu and Deshan Yang for helpful discussions about the sphaleron energy and the spectral method. This work was supported in part by the National Natural Science Foundation of China under grants No. 11835013 and No. 12235008.

\begin{appendix}
\section{Spectral Methods}
\label{app:spectral}

The EOM of the relevant fields in the calculation of the sphaleron configuration are nonlinear differential equations coupled with each other. It is usually difficult to solve them in an analytical way. In this appendix we show how to use the spectral method to numerically solve the EOM and calculate the sphaleron energy.\footnote{The code is publicly available at \href{https://github.com/Bingrong-Yu/Spectral_Sphaleron_Solver}{\color{blue}{https://github.com/Bingrong-Yu/Spectral\_Sphaleron\_Solver}}.} The main advantage of the spectral method is that it converges very quickly with high precision as the number of the grid points increases. In what follows, we first give a brief introduction to the spectral method, and then apply it to the SM and the HTM.

\subsection{Basic Ideas}
The spectral method is an efficient technique to numerically solve differential equations~\cite{Boyd,Trefethen}. The core idea is to approximate the unknown function by a set of basis functions. Let $\left\{\phi^{}_n(x)\right\}$ being a set of  orthogonal and complete functions, the unknown function $u(x)$ can be expanded as
\begin{eqnarray}
	u(x)=\sum_{n=0}^{\infty}a^{}_n \phi^{}_n(x)\;,\quad
	a^{}_n=\int {\rm d}x\, u(x) \phi^{*}_n(x)\;.
\end{eqnarray}
For practical numerical computation, one has to truncate at a finite number $n=N$, and $u(x)$ can be approximated by
\begin{eqnarray}
	\label{eq:uN}
	u(x)\approx u^{}_N(x)=\sum_{n=0}^{N}a^{}_n \phi^{}_n(x)\;,
\end{eqnarray} 
where the coefficients $a^{}_n$ are calculated at grid points $\left\{x^{}_i\right\}$
\begin{eqnarray}
	\label{eq:an}
	a^{}_n \approx \sum_{i=1}^{N}u^{}_i \phi^{*}_n(x_i)\;,
\end{eqnarray}
with $u_i\equiv u(x_i)$. Substituting Eq.~(\ref{eq:an}) back to (\ref{eq:uN}) one obtains
\begin{eqnarray}
	u^{}_N(x) = \sum_{n=0}^{N}\sum_{i=0}^{N}u^{}_i \phi_n^{*}(x_i)\phi^{}_n(x)\;.
\end{eqnarray}
Then the derivative of the unknown function can be approximated by that of the basis functions, namely
\begin{eqnarray}
	u_j' \approx u_N'(x)\Big|_{x=x_j} = \sum_{n=0}^{N}\sum_{i=0}^{N}u^{}_i \phi_n^{*}(x_i)\phi_n'(x)\Big|_{x=x^{}_j} \;.
\end{eqnarray}
The \emph{differentiation matrix} $D^{}_N$, which relates the unknown function to its derivative at grid points, is given by
\begin{eqnarray}
	\left(D_N\right)^{}_{ji}=\sum_{n=0}^{N}\phi_n^{*}(x^{}_i) \phi_n'(x)\Big|_{x=x^{}_j}\;.
\end{eqnarray}
Starting from the differentiation matrix, the values of the derivative function can be easily expressed as the linear combination of the values of the raw function. For example, we have
\begin{eqnarray}
	u_j' = \sum_{i=1}^{N} \left(D^{}_{N}\right)^{}_{ji}u^{}_i\;,\quad
	u_j'' = \sum_{i=1}^{N} \left(D_{N}^{2}\right)^{}_{ji}u^{}_{i}\;.
\end{eqnarray}
Then the differential equations of $u(x)$ are reduced to a set of algebraic equations of $\left\{u^{}_i\right\}$, which can be numerically solved directly. 

The numerical error of the above method is described by the residual function
\begin{align}
	R(x)=\left|u(x)-u^{}_N(x)\right|.
\end{align}
Therefore, a ``good choice" of the basis functions $\left\{\phi^{}_n(x)\right\}$ and the grid points $\left\{x^{}_i\right\}$ should make the residual function as small as possible. For periodic functions, the best choice of the basis functions is the Fourier series. However, for non-periodic functions, as what we encountered in the calculation of the sphaleron, it can be shown that in most cases the best choice of the basis functions is the \emph{Chebyshev polynomials} (see Appendix~\ref{app:chebyshev})~\cite{Boyd,Trefethen}. In addition, the grid points should be taken as the extrema of the Chebyshev polynomials, i.e.,
\begin{eqnarray}
	\label{eq:grid}
	x_j = \cos\left(\frac{j\pi}{N}\right)\;,\quad
	j=0,1,\cdots,N\;.
\end{eqnarray}
Then it is straightforward to construct the Chebyshev
spectral differentiation matrix~\cite{Trefethen} 
\begin{eqnarray}
	\label{eq:DN}
	\left(D^{}_{N}\right)^{}_{00} &=& \frac{2N^{2}_{}+1}{6}\;,\quad \left(D^{}_{N}\right)^{}_{NN} = -\frac{2N_{}^{2}+1}{6}\;,\nonumber\\
	\left(D^{}_{N}\right)^{}_{jj} &=& \frac{-x^{}_{j}}{2\left(1-x_{j}^{2}\right)},\quad j=1,\cdots, N-1\;,\nonumber\\
	\left(D^{}_{N}\right)^{}_{ij}&=&\frac{c^{}_{i}}{c^{}_{j}}\frac{(-1)_{}^{i+j}}{x^{}_{i}-x^{}_{j}},\quad i\neq j,\; 0\leqslant i,j \leqslant N \; ,
\end{eqnarray}
where
\begin{eqnarray}
	c^{}_{i}=\left\lbrace
	\begin{aligned}
		&2\qquad i=0\;\text{or}\; N\\
		&1\qquad \text{otherwise}\nonumber
	\end{aligned}\;.
	\right.
\end{eqnarray}
One should keep in mind that when using the Chebyshev
spectral method to solve differential equations, the following two conditions need to be satisfied
\begin{itemize}
	\item domain of the variable: $x\in [-1,1]$\; ;
	\item boundary conditions: $u(-1) = u(1) =0$\;.
\end{itemize}
They are easily to achieve after a linear transformation of the variable. In the following parts we will show how to use the spectral method introduced above to solve the differential equations relevant to the sphaleron. 

\subsection{Sphaleron in the Standard Model}
\begin{figure}[t!]
	\centering
	\includegraphics[scale=0.75]{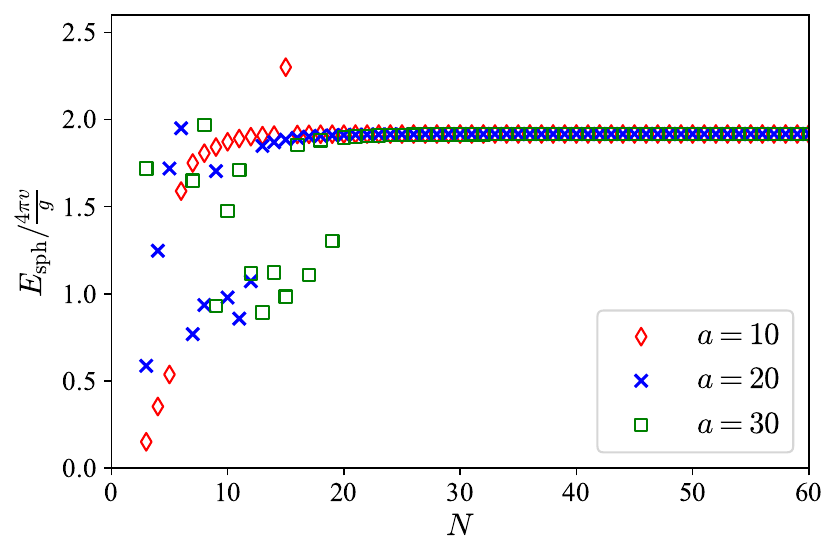}
	\caption{\label{fig:conver}The sphaleron energy in the SM obtained using the spectral method for different numbers of grid points $N$ and cut-off $a$. We have fixed $\varrho^{}_{1}=0.306$. It can be seen that the energy converges rapidly as $N$ increases, and independent of $a$ as long as $N\gtrsim 20$.}
\end{figure}
As a warm up, we first use the spectral method to calculate the sphaleron configuration in the SM. There are only two dynamical fields [i.e., $f(\xi)$ and $h(\xi)$], and their EOM are given by (recalling that we have defined $\xi\equiv g v r$ and $\varrho^{}_{1}\equiv \lambda/g^2_{}$)
\begin{eqnarray}
	\label{eq:SM-f}
	\xi_{}^2 f''&=& 2f\left(1-f\right)\left(1-2f\right)-\frac{\xi_{}^2}{4}\left(1-f\right)h_{}^2\;,\\
	\label{eq:SM-h}
	\left(\xi_{}^2 h'\right)'&=&2\left(1-f\right)_{}^2 h-\varrho_1 \xi_{}^2  h\left(1-h_{}^2\right)\;,
\end{eqnarray}
with the boundary conditions $f(0)=h(0)=0$ and $f(\infty)=h(\infty)=1$. In the practical calculation, the variable is truncated at some finite distance $\xi^{}_{\rm max}=2a$. This is reasonable because the sphaleron energy is localized near the origin and the profile functions $f$ and $h$ tend to the constant quickly as the distance increases. In order to satisfy the conditions of the Chebyshev spectral method, we perform a linear transformation to the variable
\begin{eqnarray}
	\label{eq:variabletrans}
	\xi \to x=\frac{\xi}{a}-1\;.
\end{eqnarray}
In addition, the profile functions should be shifted to
\begin{eqnarray}
	f(x) \to \bar{f}(x) = f(x) - \frac{1+x}{2}\;,\quad
	h(x) \to \bar{h}(x) = h(x) - \frac{1+x}{2}\;.
\end{eqnarray}
Then the domain of the variable is $x\in [-1,1]$ and the boundary conditions become $\bar{f}(-1)=\bar{f}(1)=\bar{h}(-1)=\bar{h}(1)=0$. The EOM of the shifted profile functions turn out to be
\begin{align}
	\label{eq:fbar}
	2\left(1+x\right)_{}^2 \bar{f}'' &= \left(2\bar{f}+1+x\right)\left(2\bar{f}-1+x\right)\left(2\bar{f}+x\right)\nonumber\\
	&+\frac{a_{}^2}{16}\left(1+x\right)_{}^2\left(2\bar{f}-1+x\right)\left(2\bar{h}+1+x\right)_{}^2\;,\\
	\label{eq:hbar}
	\left(1+x\right)_{}^2 \bar{h}'' + \left(1+x\right)\left(2\bar{h}'+1\right)	&= \frac{1}{4}\left(2\bar{f}-1+x\right)_{}^2\left(2\bar{h}+1+x\right)
	\nonumber\\
	&-\frac{ a_{}^2 \varrho^{}_{1}}{8}\left(1+x\right)_{}^2\left(2\bar{h}+1+x\right)\left[4-\left(2\bar{h}+1+x\right)_{}^2\right]\;.
\end{align}
Note that all the derivatives in Eqs.~(\ref{eq:fbar}) and (\ref{eq:hbar}) are with respective to $x$ rather than $\xi$.

Now we can use the Chebyshev spectral method introduced above to solve the EOM. Given the grid points in Eq.~(\ref{eq:grid}), it is straightforward to construct the $(N+1)\times (N+1)$ differentiation matrix $D_N$ using Eq.~(\ref{eq:DN}). The derivatives of the profile functions are given by $\bar{f}' = D^{}_N \bar{f}$, $\bar{h}' = D^{}_N \bar{h}$, $\bar{f}'' = D_N^2 \bar{f}$, and $\bar{h}'' = D_N^2 \bar{h}$. Then Eqs.~(\ref{eq:fbar}) and (\ref{eq:hbar}) are reduced to $2(N-1)$ algebraic equations with respect to 
\begin{eqnarray*}
	\left\{\bar{f}(x^{}_1),\cdots,\bar{f}(x^{}_{N-1}),\bar{h}(x^{}_1),\cdots,\bar{h}(x^{}_{N-1})\right\}\;,
\end{eqnarray*}
which can be numerically solved directly. Finally, the profile functions should be shifted back via $f(x)=\bar{f}(x)+\left(1+x\right)/2$ and $h(x)=\bar{h}(x)+\left(1+x\right)/2$, and the energy of the sphaleron could be computed by
\begin{align}
	\label{eq:SM-E}
	E^{}_{\rm sph}=\frac{4\pi v a}{g}\int_{x^{}_{N-1}}^{x^{}_1}{\rm d}x&\left\{
	\frac{4}{a^2_{}}f'^2_{}+\frac{8}{a_{}^2\left(1+x\right)_{}^2}f_{}^2\left(1-f\right)_{}^2+\left(1-f\right)_{}^2h_{}^2+\frac{1}{2}\left(1+x\right)_{}^2 h'^2_{}\right.\nonumber\\
	&\left.+\frac{\varrho^{}_{1}}{4} a_{}^2\left(1+x\right)_{}^2 \left(1-h_{}^2\right)_{}^2
	\right\}\;,
\end{align}
where the upper and lower bounds $x^{}_1$ and $x^{}_{N-1}$ are given by Eq.~(\ref{eq:grid}).

\begin{figure}[t!]
	\centering
	\includegraphics[scale=0.55]{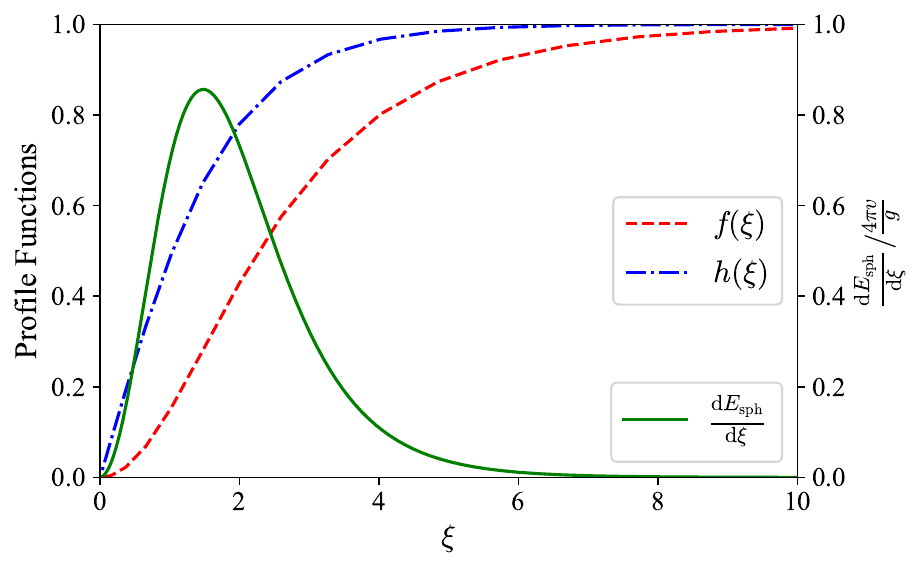}\qquad 
	\includegraphics[scale=0.55]{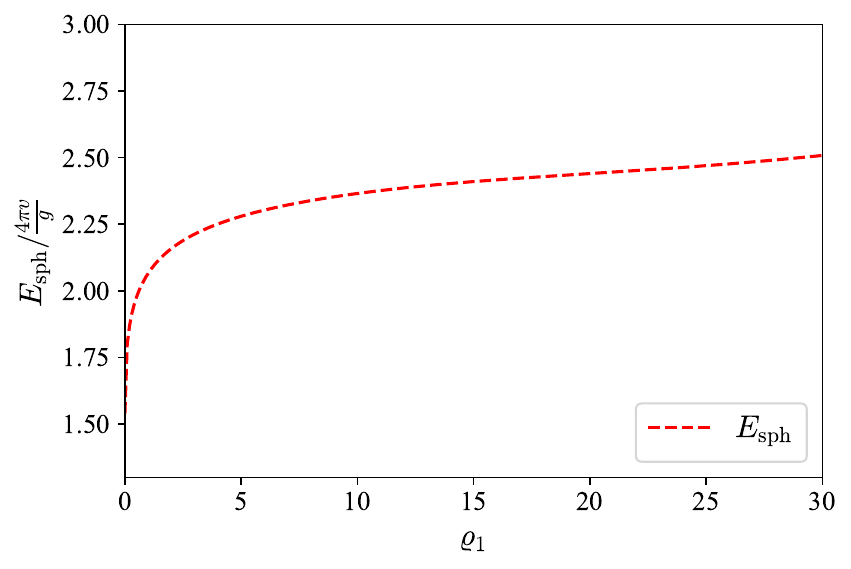}
	\caption{\label{fig:SM-Sphaleron}Sphaleron configuration in the SM obtained by the spectral method. \emph{Left}: The solutions of profile functions and the sphaleron energy density for $\varrho^{}_{1}=0.306$. \emph{Right}: The sphaleron energy versus the coupling $\varrho^{}_1$.}
\end{figure}

We find the results converge rapidly as the number of grid points $N$ increases (see Fig.~\ref{fig:conver}). For  $N \gtrsim 20$, the numerical results are stable and independent of the cut-off $a$.
This is because the profile functions and the sphaleron energy density tend to constants quickly as $\xi$ increases. In Fig.~\ref{fig:SM-Sphaleron} we show the sphaleron configuration in the SM obtained using the spectral method, where $N=60$ and $a=30$ have been taken. It is worthwhile to mention that the spectral method takes only about 1 second to calculate the sphaleron configuration for a given $\varrho^{}_{1}$ using a usual personal desktop. In particular, for $\varrho_1=0$, $\varrho^{}_{1}=0.306$ and $\varrho^{}_1 \to \infty$, we obtain $E^{}_{\rm sph} \approx 1.54$, $E^{}_{\rm sph}\approx 1.92$ and $E^{}_{\rm sph} \approx 2.71$ (in units of $4\pi v/g$), which matches very well with the result in the literature~\cite{Manton:1983nd,Klinkhamer:1984di,Yaffe:1989ms}.

\subsection{Sphaleron in the Higgs Triplet Model}
Then we turn to calculate the sphaleron configuration in the HTM using the spectral method. We have three dynamical fields, i.e., $f(\xi)$, $h(\xi)$, and $h^{}_\Delta(\xi)$. As what we did in the SM, in order to satisfy the boundary conditions of the spectral method, the variable $\xi$ should be transform to $x$ via Eq.~(\ref{eq:variabletrans}), and the profile functions should be shifted to
\begin{align}
	f(x) \to \bar{f}(x) = f(x) - \frac{1+x}{2}\;,\quad
	h(x) \to \bar{h}(x) = h(x) - \frac{1+x}{2}\;,\quad
	h^{}_\Delta(x) \to \bar{h}^{}_\Delta(x) = h^{}_\Delta(x) - \frac{1+x}{2}\;.
\end{align}
Now the domain of the variable is $x\in [-1,1]$ and the boundary conditions become $\bar{f}(-1)=\bar{f}(1)=\bar{h}(-1)=\bar{h}(1)=\bar{h}^{}_\Delta(-1)=\bar{h}^{}_\Delta(1)=0$. After some straightforward calculations, the EOM of the shifted profile functions turn out to be
{\allowdisplaybreaks
\begin{align}
	\label{eq:fbartypeII}
	2\left(1+x\right)_{}^2 \bar{f}'' &= \left(2\bar{f}+1+x\right)\left(2\bar{f}-1+x\right)\left(2\bar{f}+x\right)\nonumber\\
	&+\frac{a_{}^2}{16\beta}\left(1+x\right)_{}^2\left(2\bar{f}-1+x\right)\left(2\bar{h}+1+x\right)_{}^2\nonumber\\
	&+\frac{a_{}^2 \varrho_{3}}{6\beta}\left(1+x\right)_{}^2\left(2\bar{f}-1+x\right)\left(2\bar{h}^{}_\Delta+1+x\right)_{}^2\;,\\
	\label{eq:hbartypeII}
	\left(1+x\right)_{}^2 \bar{h}'' + \left(1+x\right)\left(2\bar{h}'+1\right)&= \frac{1}{4}\left(2\bar{f}-1+x\right)_{}^2\left(2\bar{h}+1+x\right)\nonumber\\
	&-\frac{a_{}^2}{8\beta}\left(1+x\right)_{}^2\left\{\left(\varrho^{}_{1}-\varrho^{}_{2}\right)\left(2\bar{h}+1+x\right) \left[4-\left(2\bar{h}+1+x\right)_{}^2\right]\right.\nonumber\\
	&\left.-\varrho^{}_{2}\left(2\bar{h}+1+x\right)\left[\left(2\bar{h}+1+x\right)_{}^2-2\left(2\bar{h}^{}_\Delta+1+x\right)\right]\right.\nonumber\\
	&\left.+4\left(\varrho^{}_{4}-\varrho^{}_{1}+\varrho^{}_{2}\right)\left(2\bar{h}+1+x\right)\right.\nonumber\\
	&\left.+2\left(\sqrt{2\varrho^{}_{2}\varrho^{}_{3}\varrho^{}_{5}}-\varrho^{}_{2}\right)\left(2\bar{h}+1+x\right)\left(2\bar{h}^{}_\Delta+1+x\right)\right.\nonumber\\
	&\left.+\left(\varrho^{}_{1}-\varrho^{}_{4}-\sqrt{2\varrho^{}_{2}\varrho^{}_{3}\varrho^{}_{5}}\right)\left(2\bar{h}+1+x\right)\left(2\bar{h}^{}_\Delta+1+x\right)_{}^2\right\}\;,\\
	\label{eq:hbarDeltatypeII}
	\varrho^{}_{3} \left(1+x\right)_{}^2 \bar{h}_\Delta'' + \varrho^{}_{3} \left(1+x\right)\left(2\bar{h}_\Delta'+1\right)	&= \frac{2\varrho^{}_{3} }{3}\left(2\bar{f}-1+x\right)_{}^2\left(2\bar{h}^{}_\Delta+1+x\right)\nonumber\\
	&-\frac{a_{}^2 \varrho^{}_{2}}{8\beta} \left(1+x\right)_{}^2\left[\left(2\bar{h}+1+x\right)_{}^2-2\left(2\bar{h}^{}_\Delta+1+x\right)\right]\nonumber\\
	&+\frac{a_{}^2}{8\beta}\left(1+x\right)_{}^2\left[2\left(2\varrho^{}_{3}\varrho^{}_{5}-\varrho^{}_{2}\right)\left(2\bar{h}^{}_\Delta+1+x\right)\right.\nonumber\\
	&\left.-\left(\sqrt{2\varrho^{}_{2}\varrho^{}_{3}\varrho^{}_{5}}-\varrho^{}_{2}\right)\left(2\bar{h}+1+x\right)^2\right.\nonumber\\
	&\left.-\left(\varrho^{}_{1}-\varrho^{}_{4}-\sqrt{2\varrho^{}_{2}\varrho^{}_{3}\varrho^{}_{5}}\right)\left(2\bar{h}+1+x\right)_{}^2
	\left(2\bar{h}^{}_\Delta+1+x\right)\right.\nonumber\\
	&\left.+\left(\varrho^{}_{1}-\varrho^{}_{4}-\varrho^{}_{3}\varrho^{}_{5}-\sqrt{\varrho^{}_{2}\varrho^{}_{3}\varrho^{}_{5}/2}\right)\left(2\bar{h}^{}_\Delta+1+x\right)_{}^3\right]\;.
\end{align}
}
Note that all the derivatives are with respective to $x$.
If we take $\varrho^{}_{4}=\varrho^{}_{1}-\varrho^{}_{2}$ and $\varrho^{}_{5}=\varrho^{}_{2}/\left(2\varrho^{}_{3}\right)$, then Eqs.~(\ref{eq:fbartypeII})-(\ref{eq:hbarDeltatypeII}) simply reduce to the EOM of the shifted profile functions in the minimal HTM.
Constructing the differentiation matrix $D_N$ using Eq.~(\ref{eq:DN}), the derivatives of the profile functions are given by 
\begin{eqnarray}
	\bar{f}' &=& D_N \bar{f}\;,\quad  \bar{h}' = D_N \bar{h}\;,\quad 
	\bar{h}_\Delta' = D_N \bar{h}_\Delta\;,\nonumber\\
	\bar{f}'' &=& D_N^2 \bar{f}\;,\quad  \bar{h}'' = D_N^2 \bar{h}\;,\quad 
	\bar{h}''_\Delta= D_N^2 \bar{h}^{}_\Delta\;.
\end{eqnarray}
Then Eqs.~(\ref{eq:fbartypeII})-(\ref{eq:hbarDeltatypeII}) reduce to $3(N-1)$ algebraic equations with respective to 
\begin{eqnarray*}
	\left\{\bar{f}(x^{}_1),\cdots,\bar{f}(x^{}_{N-1}),\bar{h}(x^{}_1),\cdots,\bar{h}(x^{}_{N-1}),\bar{h}^{}_\Delta(x^{}_1),\cdots,\bar{h}^{}_\Delta(x^{}_{N-1})\right\}\;,
\end{eqnarray*}
and they can be numerically solved directly. The profile functions should be shifted back: $f(x)=\bar{f}(x)+\left(1+x\right)/2$, $h(x)=\bar{h}(x)+\left(1+x\right)/2$, and $h^{}_\Delta(x)=\bar{h}^{}_\Delta(x)+\left(1+x\right)/2$. Finally, the energy of the sphaleron is calculated by
{\allowdisplaybreaks
\begin{align}
	E_{\rm sph}=\frac{4\pi v a}{g}\int_{x^{}_{N-1}}^{x^{}_1}{\rm d}x & \left\{
	\frac{4}{a_{}^2}f'^2_{}+\frac{8}{a_{}^2\left(1+x\right)_{}^2}f_{}^2\left(1-f\right)_{}^2+\frac{1}{\beta}\left(1-f\right)^2h_{}^2+\frac{1}{2\beta}\left(1+x\right)_{}^2 h'^2_{}\right.\nonumber\\
	&\left.+\frac{a_{}^2\left(1+x\right)_{}^2}{4\beta^2_{}}\left[\left(\varrho^{}_1-\varrho^{}_2\right)\left(1-h_{}^2\right)^2_{}+\varrho^{}_2\left(h_{}^2-h^{}_\Delta\right)^2_{}\right]+\frac{\varrho^{}_3}{6\beta}\left[3\left(1+x\right)_{}^2 h_\Delta'^2\right.\right.\nonumber\\
	&\left.\left.
	+16h_\Delta^2 \left(1-f\right)_{}^2\right]+\frac{a_{}^2\left(1+x\right)^2}{4\beta^2_{}}\left[2\left(\varrho^{}_4-\varrho^{}_1+\varrho^{}_2 \right)\left(1-h_{}^2\right)\right.\right.\nonumber\\
	&\left.\left.-\left(2\varrho^{}_3 \varrho^{}_5 -\varrho^{}_2\right)\left(1-h_\Delta^2\right)\right]+\frac{a_{}^2\left(1+x\right)_{}^2}{2\beta_{}^2}\left(\sqrt{2\varrho^{}_2 \varrho^{}_3 \varrho^{}_5}-\varrho^{}_2\right)\left(1-h_{}^2 h^{}_\Delta\right)\right.\nonumber\\
	&\left.+\frac{a_{}^2\left(1+x\right)_{}^2}{2\beta_{}^2}\left(\varrho^{}_1-\varrho^{}_4-\sqrt{2\varrho^{}_2 \varrho^{}_3 \varrho^{}_5}\right)\left(1-h_{}^2 h_\Delta^2\right)\right.\nonumber\\
	&\left.-\frac{a^2_{}\left(1+x\right)^2_{}}{4\beta^2_{}}\left(\varrho^{}_1-\varrho^{}_4-\varrho^{}_3 \varrho^{}_5-\sqrt{\varrho^{}_2 \varrho^{}_3 \varrho^{}_5/2}\right)\left(1-h_\Delta^4\right)\right\}\;,
\end{align}
}
where $x^{}_1 = \cos\left(\pi/N\right)$ and $x^{}_{N-1}=\cos\left[\left(N-1\right)\pi/N\right]=-\cos\left(\pi/N\right)$. As in the SM, we find the final results converge rapidly as $N$ increases and depend very weakly on $a$. Therefore, in the numerical calculation throughout this work, we fix $N=60$ and $a=30$.

\section{Chebyshev Polynomials}
\label{app:chebyshev}
In this mathematical appendix, we briefly review some properties of the Chebyshev polynomials. We also demonstrate why the Chebyshev polynomials serve as a ``good candidate" of the basis functions in the spectral method. 

The Chebyshev polynomial of degree $n$ is defined as 
\begin{eqnarray}
	T^{}_n\left(\cos\theta\right)=\cos\left(n\theta\right)\;,\quad
	n=0,1,2,\cdots\;.
\end{eqnarray}
From the definition one can obtain
\begin{eqnarray}
	T^{}_{0}(x) = 1\;,\quad
	T^{}_{1}(x) = x\;,\quad
	T^{}_{n+2}(x) = 2xT^{}_{n+1}(x)-T^{}_{n}(x)\;.
\end{eqnarray}
It is easy to show that the Chebyshev polynomials satisfy the following properties:
\begin{itemize}
	\item Orthonormality. The Chebyshev polynomials are orthogonal with respect to the weight function $\rho(x)=1/\sqrt{1-x_{}^2}$, i.e., 
	\begin{align}
		&\int_{-1}^1 \frac{{\rm d}x}{\sqrt{1-x^2}}T^{}_m(x) T^{}_n(x) = 0 \qquad \text{for}\;m\neq n\;,\nonumber\\
		&\int_{-1}^{1} \frac{{\rm d}x}{\sqrt{1-x^2}}T_n^2(x) = 
		\left\lbrace	
		\begin{aligned}
			&\pi\qquad   {\rm for}\; n=0\\
			&\pi/2\quad {\rm for}\; n=1,2,3,\cdots
		\end{aligned}\;.
		\right.
	\end{align}
	\item Completeness. Any function $u(x)$ defined on $[-1,1]$ can be expanded as 
	\begin{eqnarray}
		\label{eq:ChebyshevExpand}
		u(x) = \sideset{}{'} \sum_{n=0}^{\infty} a^{}_n T^{}_n(x)\;,\qquad
		a^{}_n=\frac{2}{\pi}\int_{-1}^{1}\frac{{\rm d}x}{\sqrt{1-x^2}}u(x)T^{}_n(x)\;,
	\end{eqnarray}
	where $\sum'$ denotes a sum whose first term is halved.
	
	\item Roots and extrema. The Chebyshev polynomial of degree $n$ has $n+1$ extrema and $n$ roots  in $[-1,1]$
	\begin{eqnarray}
		\label{eq:extrema}
		{\rm extrema}: \quad x^{}_j &=& \cos\left(\frac{j\pi}{n}\right)\;, \quad j=0,1,\cdots,n\;,\\
		\label{eq:roots}
		{\rm roots}: \quad \tilde{x}^{}_j &=& \cos\left(\frac{2j+1}{2n}\pi\right)\;, \quad j=0,1,\cdots,n-1\;.
	\end{eqnarray}
\end{itemize}

For the practical numerical calculation, the infinite sum in Eq.~(\ref{eq:ChebyshevExpand}) should be truncated at $n=N$, and the coefficients are evaluated at grid points~\cite{Elliott1965,1969Chebyshev}
\begin{eqnarray}
	\label{eq:extremagrid}
	u(x)\approx u^{}_N(x) = \sideset{}{''} \sum_{j=0}^{N} b^{}_n T^{}_n(x)\;,\qquad b^{}_n = \frac{2}{N} \sideset{}{''} \sum_{n=0}^{N} u(x^{}_j) T^{}_n(x^{}_j)\;,
\end{eqnarray}
where $\sum''$ denotes a sum whose first and last terms are halved, and $x^{}_j = \cos\left(j\pi/N\right)$ (for $j=0,1,\cdots,N$) are extrema of the Chebyshev polynomial of degree $N$. Alternatively, one can also evaluate the coefficients at roots of the Chebyshev polynomials
\begin{eqnarray}
	\label{eq:rootsgrid}
	u(x)\approx \tilde{u}^{}_N(x)= \sideset{}{'} \sum_{n=0}^{N} \tilde{b}^{}_n T^{}_n(x)\;,\qquad 
	\tilde{b}^{}_n = \frac{2}{N+1}\sum_{j=0}^{N} u(\tilde{x}^{}_j) T^{}_n(\tilde{x}^{}_j)\;,
\end{eqnarray}
where $\tilde{x}^{}_j=\cos\left[\left(2j+1\right)\pi/\left(2N+2\right)\right]$ (for $j=0,1,\cdots,N$) are roots of of the Chebyshev polynomial of degree $N+1$. Then it follows that the interpolation functions $u^{}_N(x)$ and $\tilde{u}^{}_N(x)$ fit $u(x)$ exactly at the grid points, i.e.,  $u^{}_N(x^{}_j)=u(x^{}_j)$ and $\tilde{u}^{}_N(\tilde{x}^{}_j)=u(\tilde{x}^{}_j)$. Moreover, it can be proved that the upper bounds of the residue functions turn out to be~\cite{Elliott1965,1969Chebyshev}
\begin{eqnarray}
	\left|u(x)-u^{}_N(x)\right| \leqslant 2 \sum_{n=N+1}^{\infty} \left|a^{}_n\right|\;,\\
	\left|u(x)-\tilde{u}^{}_N(x)\right| \leqslant 2 \sum_{n=N+1}^{\infty} \left|a^{}_n\right|\;.
\end{eqnarray} 
This means the error of evaluating the coefficients at grid points can never exceed twice the error of computing the coefficients using the integral in  Eq.~(\ref{eq:ChebyshevExpand}). The grid points in Eqs.~(\ref{eq:extremagrid}) and (\ref{eq:rootsgrid}) are known as \emph{extrema grid} and \emph{roots grid}, respectively. Both of them have been widely used in the Chebyshev spectral method~\cite{Boyd}. In this work, we take the grid points to be extrema grid [see Eq.~(\ref{eq:grid})].

One could also compare the interpolation using the Chebyshev polynomials with other polynomials. First, recall that the general Lagrange interpolation of $u(x)$ is given by
\begin{eqnarray}
	L(x) = \sum_{j=0}^{N}u^{}_j \ell^{}_j(x)\;,
\end{eqnarray}
where $u^{}_j\equiv u(x^{}_j)$ and
\begin{eqnarray}
	\ell^{}_{j}(x)=\frac{1}{c^{}_{j}}\prod^{N}_{\substack{k=0\\ k\neq j}}(x-x^{}_{k})\;,\qquad 
	c^{}_{j}=\prod^{N}_{\substack{k=0\\ k\neq j}}(x^{}_{j}-x^{}_{k})\; .
\end{eqnarray}
Then we have $L(x^{}_j)=u(x^{}_j)$ (for $j=0,1,\cdots N$). The remainder of the Lagrange interpolation reads
\begin{eqnarray}
	R(x)=u(x)-L(x)=\frac{u_{}^{(N+1)}(\zeta)}{\left(N+1\right)!}P^{}_{N+1}(x)\;,\quad P^{}_{N+1}(x)\equiv \left(x-x^{}_1\right)\cdots\left(x-x^{}_N\right)\;,
\end{eqnarray}
where $u_{}^{(N)}(x)$ is the $N$-th derivative of $u(x)$ and $\zeta \in (-1,1)$. The question is: how to choose the grid points $x_j$ so that we could have the smallest remainder? An intuitive answer is to look at the upper bound of the remainder, which turns out to be
\begin{eqnarray}
	{\rm max}\left|R(x)\right|\leqslant \frac{{\rm max}\left|u_{}^{(N+1)}(x)\right|}{(N+1)!}\,{\rm max}\left|P^{}_{N+1}(x)\right|\;.
\end{eqnarray}
It is not difficult to prove that
\begin{eqnarray}
	{\rm max}\left|P^{}_{N+1}(x)\right|\geqslant \frac{1}{2^N_{}} {\rm max}\left|T^{}_{N+1}(x)\right|=\frac{1}{2^N_{}}\;.
\end{eqnarray}
If $P^{}_{N+1}(x)$ is the monic Chebyshev polynomial $T^{}_{N+1}(x)/2^N$, namely the grid points $x_j$ are taken to be the roots of $T^{}_{N+1}(x)$, then ${\rm max}\left|R(x)\right|$ has the minimum upper bound. Therefore, the Chebyshev polynomial is the ``best choice" of the interpolation polynomial, in the sense that the remainder has a minimum upper bound.

\end{appendix}
\bibliographystyle{elsarticle-num}
\bibliography{ref}
\end{document}